\documentclass[10pt,journal,compsoc]{IEEEtran}
\ifCLASSOPTIONcompsoc
  \usepackage[nocompress]{cite}
\else
  \usepackage{cite}
\fi

\ifCLASSINFOpdf
 
\else

\fi

\usepackage{mathptmx}
\usepackage{times}

\usepackage{array}
\usepackage{hyperref}
\usepackage{wrapfig}
\usepackage{amsmath}
\usepackage{amssymb}
\usepackage{tikz}
\usetikzlibrary{shapes,fit,arrows,positioning,shapes.geometric}
\usepackage{algorithm}
\usepackage{algpseudocode}
\usepackage{subcaption}
\usepackage{amssymb}% http://ctan.org/pkg/amssymb
\usepackage{pifont}% http://ctan.org/pkg/pifont

\newcounter{bigfig}
\newcommand{\refbigfig}[1]{\refstepcounter{bigfig}\label{#1}}

\usepackage[percent]{overpic}

\algnewcommand{\LineComment}[1]{\State // {\it #1}}

\algnewcommand{\Break}{\textbf{break}}
\algtext*{EndWhile}% Remove "end while" text
\algtext*{EndIf}% Remove "end if" text
\algtext*{EndFor}% Remove "end for" text
\algtext*{EndFunction}% Remove "end function" text

\algnewcommand{\Input}[1]{\State\textbf{input: }#1}
\algnewcommand{\Output}[1]{\State\textbf{output: }#1}

\newcolumntype{C}[1]{>{\centering\let\newline\\\arraybackslash\hspace{0pt}}m{#1}}

\begin{document}

\title{LSQT: Low-Stretch Quasi-Trees\\ for Bundling and Layout}
%

%by academic/faculty status and alphabetical order for now
\author{Rebecca~Vandenberg$^\dagger$,
        Madison~Elliott$^\dagger$,
        Nicholas~Harvey,
        and~Tamara~Munzner \IEEEmembership{Senior Member, IEEE}% <-this % stops a space
        
% \IEEEcompsocitemizethanks{\IEEEcompsocthanksitem M. Shell was with the Department
% of Electrical and Computer Engineering, Georgia Institute of Technology, Atlanta,
% GA, 30332.\protect\\
% % note need leading \protect in front of \\ to get a newline within \thanks as
% % \\ is fragile and will error, could use \hfil\break instead.
% E-mail: see http://www.michaelshell.org/contact.html
% \IEEEcompsocthanksitem J. Doe and J. Doe are with Anonymous University.}% <-this % stops an unwanted space
\IEEEcompsocitemizethanks{\IEEEcompsocthanksitem  Rebecca Vandenberg (McKnight) ($^\dagger$joint first author) was with the University of British Columbia Department of Computer Science and now is with Amazon. Email: rebmck@amazon.com.
\IEEEcompsocthanksitem  Madison Elliott ($^\dagger$joint first author) is with the University of Columbia Department of Psychology, Email: mellio10@psych.ubc.ca. 
\IEEEcompsocthanksitem Nick Harvey and Tamara Munzner are with University of British Columbia Department of Computer Science. E-mail: nickhar@cs.ubc.ca, tmm@cs.ubc.ca. 
}% <-this % stops an unwanted space

\thanks{Manuscript received XXX; revised XXX.}}

\IEEEtitleabstractindextext{%
\begin{abstract}
We introduce low-stretch trees to the visualization community with LSQT, our novel technique that uses quasi-trees for both layout and edge bundling. Our method offers strong computational speed and complexity guarantees by leveraging the convenient properties of low-stretch trees, which accurately reflect the topological structure of arbitrary graphs with superior fidelity compared to arbitrary spanning trees. Low-stretch quasi-trees also have provable sparseness guarantees, providing algorithmic support for aggressive de-cluttering of hairball graphs. LSQT does not rely on previously computed vertex positions and computes bundles based on topological structure before any geometric layout occurs. Edge bundles are computed efficiently and stored in an explicit data structure that supports sophisticated visual encoding and interaction techniques, including dynamic layout adjustment and interactive bundle querying. Our unoptimized implementation handles graphs of over 100,000 edges in eight seconds, providing substantially higher performance than previous approaches. 
\end{abstract}

}

\maketitle

\IEEEdisplaynontitleabstractindextext

\IEEEpeerreviewmaketitle

\IEEEraisesectionheading{\section{Introduction}\label{sec:introduction}}

\IEEEPARstart{G}{raph} visualization poses many challenges, most of which become increasingly problematic as a graph's size increases~\cite{vonlandesberger2011visual}. 
One of the greatest challenges in graph visualization occurs when a large graph display is too cluttered to interpret or explore. Two main approaches for taming this graph complexity problem, the so-called \textit{hairball} problem, exist in the current visualization literature: first, developing novel layouts to find better spatial positions for nodes and edges; second, developing novel edge-bundling approaches to group edges together and reduce visual clutter. These approaches have typically been approached separately. Here, we introduce a technique that addresses both of them.

In this work, we revisit the idea of quasi-trees, which had previously been used only for graph layout, by showing their utility for edge bundling. Intuitively, a quasi-tree is any tree extracted from a graph that is used as a representative proxy for the full graph. For example, one previous quasi-tree approach uses a backbone of spanning tree edges as the skeleton for a graph layout, where node positions are determined by a tree layout algorithm and the edges included in the tree are straightforward to draw~\cite{munzner1997h3}. The non-tree edges, called remainder edges, are excluded from the layout process that determines node positions but can be routed according to the backbone of the tree, and can then be drawn or hidden on demand. Our new idea is to use these remainder edges to explicitly compute a data structure of bundles.
% supporting more sophisticated visual encoding and interaction techniques than could be proposed with previous approaches that relegate bundles to implicit status. 

We accomplish these advances through the use of low-stretch trees, a mathematical formalism that exposes underlying hierarchical structure in relational datasets~\cite{AKPW95}. Importantly, low-stretch trees can be used to find tree structure in any graph, even those that do not have apparently tree-like characteristics. We propose low-stretch trees as a useful tool for the graph visualization community, and use them to help expand the reach of quasi-tree methods beyond their previously limited scope of obviously tree-like sparse graphs. 

We offer two contributions to the graph visualization literature. Our first contribution is the introduction of low-stretch trees for graph visualization, which allows us to expand the scope of quasi-tree methods beyond their previously assumed limits.  Our second contribution is the novel LSQT algorithm for bundling and layout using low-stretch quasi-trees. The bundles are explicitly computed from the graph topology before any geometric layout occurs. This algorithm segments remainder edges into bundles using efficient queries on paths within the quasi-tree, leading to algorithmic speed and quality guarantees because the quasi-trees that it computes are low-stretch. We demonstrate an implementation of this algorithm in a simple proof-of-concept interactive viewer. 
%Our third contribution is to discuss the advantages of explicitly computing bundles as first-class data abstractions, using that data structure to enable novel interaction and encoding idioms that we demonstrate in a simple proof-of-concept interactive viewer.
%Our fourth contribution is an implementation of LSQT and a simple proof-of-concept interactive viewer that uses it to demonstrate the potential of encoding and interaction with explicitly computed bundles.

\section{Quasi-Tree Rationale}

We provide background on quasi-tree approaches to layout and discuss the implications of extending these ideas to edge bundling. 
  
\subsection{Layout}

Four major families of layout methods dominate the previous graph visualization literature: force-directed \cite{holten2009force}, geographic~\cite{BoutsThesis17}, cluster-based \cite{fox1997cluster, cadez2008cluster}, and adjacency matrix \cite{henry2006nodetrix} views. A fifth family of layout methods, quasi-trees, does exist but is currently under-appreciated, with only a few examples in the existing literature such as H3~\cite{munzner1997h3}, SPF~\cite{archambault2006smashing}, and the focus-based filtering approach of Boutin et al.~\cite{boutin2006focus}.  %The H3 layout was an early approach showing the utility of quasi-trees for graph layout~\cite{munzner1997h3}.  
Quasi-tree layout methods extract a hierarchical structure, specifically a spanning tree, from a general graph. Every graph node is a tree node and edges are split into two classes: \textbf{backbone} edges within the spanning tree ($E_T$), which are used to drive the node layout, and \textbf{remainder} edges ($E_R$), which are all edges not in the tree. Formally, for a graph $G=(V,E)$, the remainder edges $E_R$ = $E \ E_T$. 
For example, a previous two-phase  approach~\cite{munzner1997h3} first uses standard tree layout algorithms to lay out all of the nodes and the backbone links, and then routes the remainder edges in a second pass using the computed node positions. An alternative is to use standard force-directed placement for the entire graph~\cite{boutin2006focus}. 

The remainder edges will typically cause significant visual clutter, since the resulting layout is not optimized to avoid those edges crossing the backbone edges or the node positions. The simple clutter mitigation strategies proposed in previous work include interaction techniques where remainder edges are drawn only on demand for the edges incident to a selected node or subtree, and visual encoding techniques where remainder edges are drawn differently from backbone edges~\cite{munzner1997h3}. 

%The crucial problem is to characterize when a graph's true structure can be adequately captured by any kind of hierarchical backbone; using a quasi-tree layout would be misleading for users if the imposed hierarchical layout is a mismatch for the true graph structure. 

Using a quasi-tree layout would be misleading for users if the imposed hierarchical layout does not adequately capture the graph's true structure, because the clutter reduction strategies would hide important information about that structure rather than secondary details. Clearly quasi-tree layouts are suitable when graphs have near-tree, sparse structure leading to only a small proportion of remainder edges compared to backbone edges. Past work by Archambault et al.~\cite{archambault2006smashing} does propose quasi-tree layouts for de-cluttering dense graphs in a ``last-ditch'' context where other layout approaches result in uninterpretable hairball problems. 
%Past work from Boutin et al.~\cite{boutin2006focus} and Munzner \cite{munzner1997h3} shows that \\ \tm{now that i think about it we did not show that in H3. there were some very dense graphs, we just filtered out most of their structure!! todo check if boutin had sparse or dense base graphs} 
Inspired by the potential of quasi-trees to tame visual clutter, we argue that these methods are a viable first choice when we use low-stretch trees because they capture true structure even in graphs that are not obviously trees. Our approach shows the suitability of quasi-tree methods for de-cluttering a much larger class of graphs than previously understood. 

\subsection{Bundling}

Beyond layout, we argue that quasi-tree methods are viable approaches to taming graph complexity with edge bundling. We propose a novel bundling system, where backbone edges of a tree are not bundled. Instead, only the remainder edges are bundled, and we explicitly compute these bundles in order to store them in a data structure that can be exploited by visual encoding and interaction techniques. We subdivide the remainder edges into segments, which can then be routed according to the existing backbone edge structure. Segments with shared endpoints are collected into bundles.  We adopt the useful terminology of \textbf{routing graph} from Bouts and Speckmann~\cite{BS15}, meaning the scaffolding for determining the position of the remainder edges. 

% There are many advantages to explicitly computing bundles. By promoting bundles to a first-class data abstraction in the visualization process, this data structure can be harnessed for the fundamental goal of edge-bundling methods, namely showing coarse-scale connectivity when fine-scale structure is too dense to interpret \cite{lee2006task, zhou2013edge}. For example, the LSQT algorithm enables an interface with three categories of edges: backbone edges, remainder edges, and edge bundles.
% %Users can distinguishably draw these categories, and manipulate their look and feel in the display independently from one another. 
% These three types of edges can be distinguishably visually encoded and interacted with independently from each other. The users can reduce clutter by drawing aggregate bundles but not remainder edges, or choose to display multi-level structure by simultaneously drawing both bundles and remainder edges, or drill down to see details for a region of interest by selecting a bundle to see all of its underlying remainder edges highlighted along with their node endpoints. 
% %\me{this will be impacted by SBEB paper. TM: addressed} 
% These interaction and encoding idioms yield obvious benefits to users exploring a graph display, but are not supported in previous edge-bundling work because nearly all existing methods use implicit, rather than explicit, bundling approaches. 
% %The ability to draw explicit bundles, in addition to their constituent remainder edges, is a novel approach to edge bundling for graph visualization.  

Another implication of our quasi-tree approach to bundling is that it allows the bundling computation to occur before any geometric layout, because it is based only on the graph's topology. 
%The geometric layout of the quasi-tree can be handled by any existing tree layout method.
Our approach allows complete independence from the layout when bundling: no initial geometric layout is required. This approach is particularly suitable when a graph layout has not already been computed, or an existing graph layout is either unintuitive (for example, a geographic layout that is a mismatch for a topologically focused task), or uninterpretable (for example, a hairball where the geometry does not capture the topological detail). It is important to note that our method is unsuitable if there is an existing graph layout that captures useful structure in the data given an intended task. In this case, previous methods that strive to preserve the original layout \cite{kumar2008system, becker1995visualizing, phan2005flow} are a better choice. 

%We also identified several advantages of quasi-tree bundling for taming complexity in graph displays. 
Just as quasi-tree approaches to layout strongly prioritize clutter removal, quasi-tree bundling approaches are an aggressive approach to de-cluttering. They are similar in spirit to ink minimization edge bundling approaches \cite{hurter2012graph, gansner2011multilevel}, which prioritize de-cluttering. In contrast, other families of approaches such as image-based or geometry-based bundling \cite{gansner2011multilevel, BS15, PNK11} have the goal of preserving the original graph layout. 

\section{Low-Stretch Quasi-Trees}

We define low-stretch trees and explain their properties, discuss the algorithm for extracting them from a graph and its complexity, and present an analysis of their suitability for our purposes.  

\subsection{Definition and Guarantees}
\label{sec:lst}

Low-stretch trees have been developed in the theoretical computer science community over the past twenty five years. They are related to the notion of spanners, which have been studied in the graph theory community for many decades \cite{AKPW95}. A spanner can be defined as a subgraph that approximately preserves edge distances. Given a graph and a spanning tree, the stretch \cite{AKPW95} is the ratio between path length in the tree and path length in the original graph. A low-stretch tree is a spanning tree that approximately minimizes the stretch of edges on average. Such a tree provides a desirable preservation of distance for the creation of a good routing graph that is used to lay out the remainder edges. These trees are the key objects underlying several recent breakthroughs in spectral graph algorithms, such as maximum flow in near-linear time \cite{LXBbook}, but they have yet to be exploited for graph visualization. 

\begin{figure}[h]
\captionsetup{justification=raggedright}
\centering
          \begin{subfigure}[t]{0.45\columnwidth}
       \centering
                \includegraphics[width=0.9\textwidth]{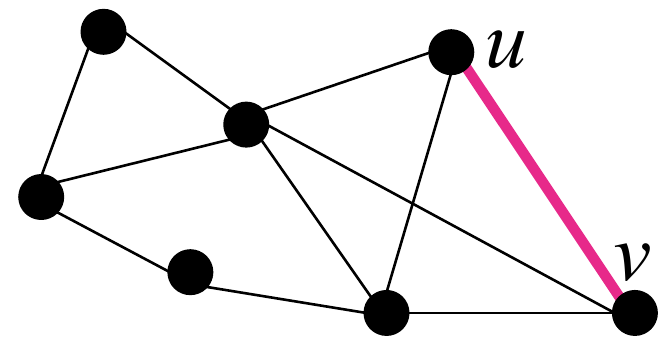}
                \caption{Original graph $G$, with comparison edge $e=(u,v)$ highlighted}
                \label{fig:stretch:a}
        \end{subfigure}
        ~
        \begin{subfigure}[t]{0.45\columnwidth}
       \centering
                \includegraphics[width=0.9\textwidth]{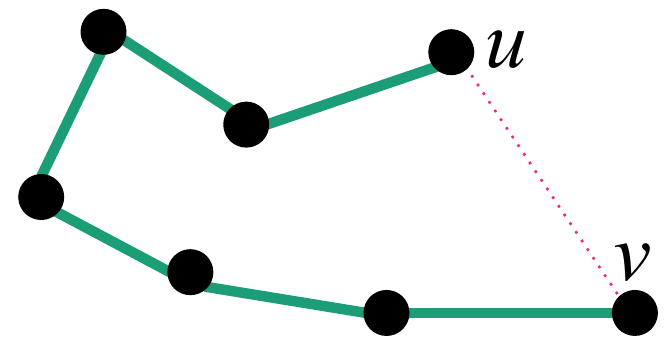}
                \caption{Tree $T$ with $s_T(e)= 6$, $s_T(G)= {28}/{11}$}
                \label{fig:stretch:b}
        \end{subfigure}
        \begin{subfigure}[t]{0.45\columnwidth}
       \centering
                \includegraphics[width=0.9\textwidth]{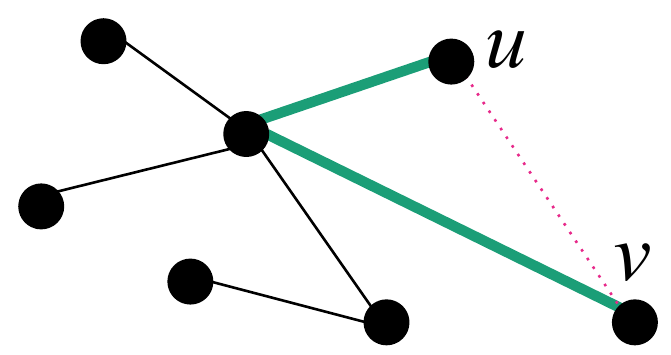}
                \caption{Tree $T$ with $s_T(e)= 2$, $s_T(G)={17}/{11}$}
                \label{fig:stretch:c}
        \end{subfigure}
  \vspace{-5pt}
  \caption{Comparison between trees with poor (high) and good (low) stretch, respectively.}
  \label{fig:stretch}
\end{figure}

\autoref{fig:stretch} gives an example of two spanning trees: one with poor (large) stretch, and one with good (small) stretch.
For graph $G=(V,E)$, the stretch of an edge $e = (u,v) \in E$ in tree $T$ is defined as:
$$s_T(e) = d_T(u,v)$$
where $d_T(u,v)$ is the path length from $u$ to $v$ in $T$. The overall stretch of $G$ is defined as:
$$s_T(G) = \frac{1}{|E|}\sum_{e\in E} s_T(e)$$

\begin{figure}[h]
	\centering
	\begin{subfigure}[t]{0.45\columnwidth}
       \centering
                \includegraphics[width=.9\textwidth]{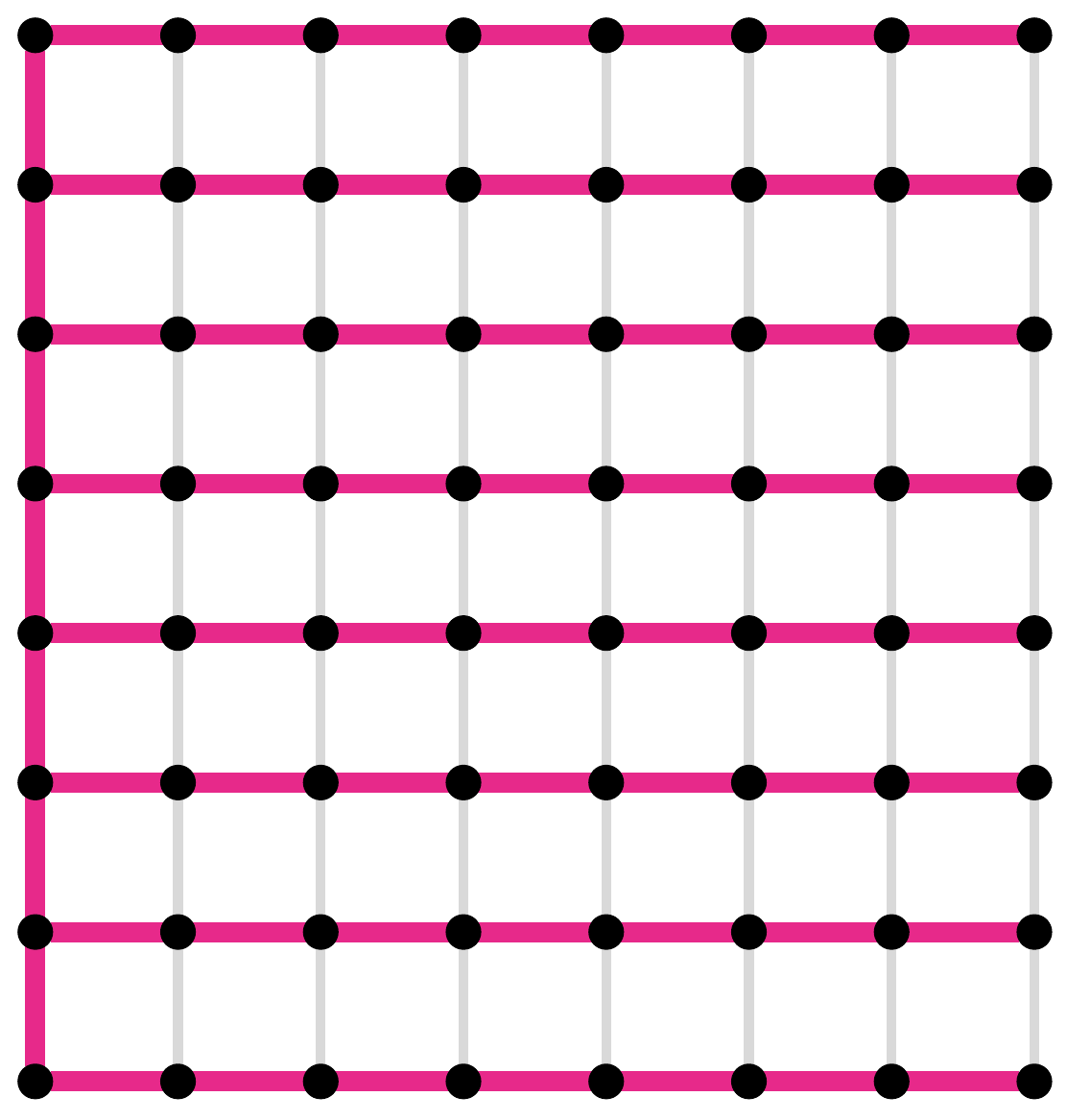}
                \caption{A spanning tree.}
                \label{fig:grid-mst}
        \end{subfigure}
        ~
        \begin{subfigure}[t]{0.45\columnwidth}
       \centering
                \includegraphics[width=.9\textwidth]{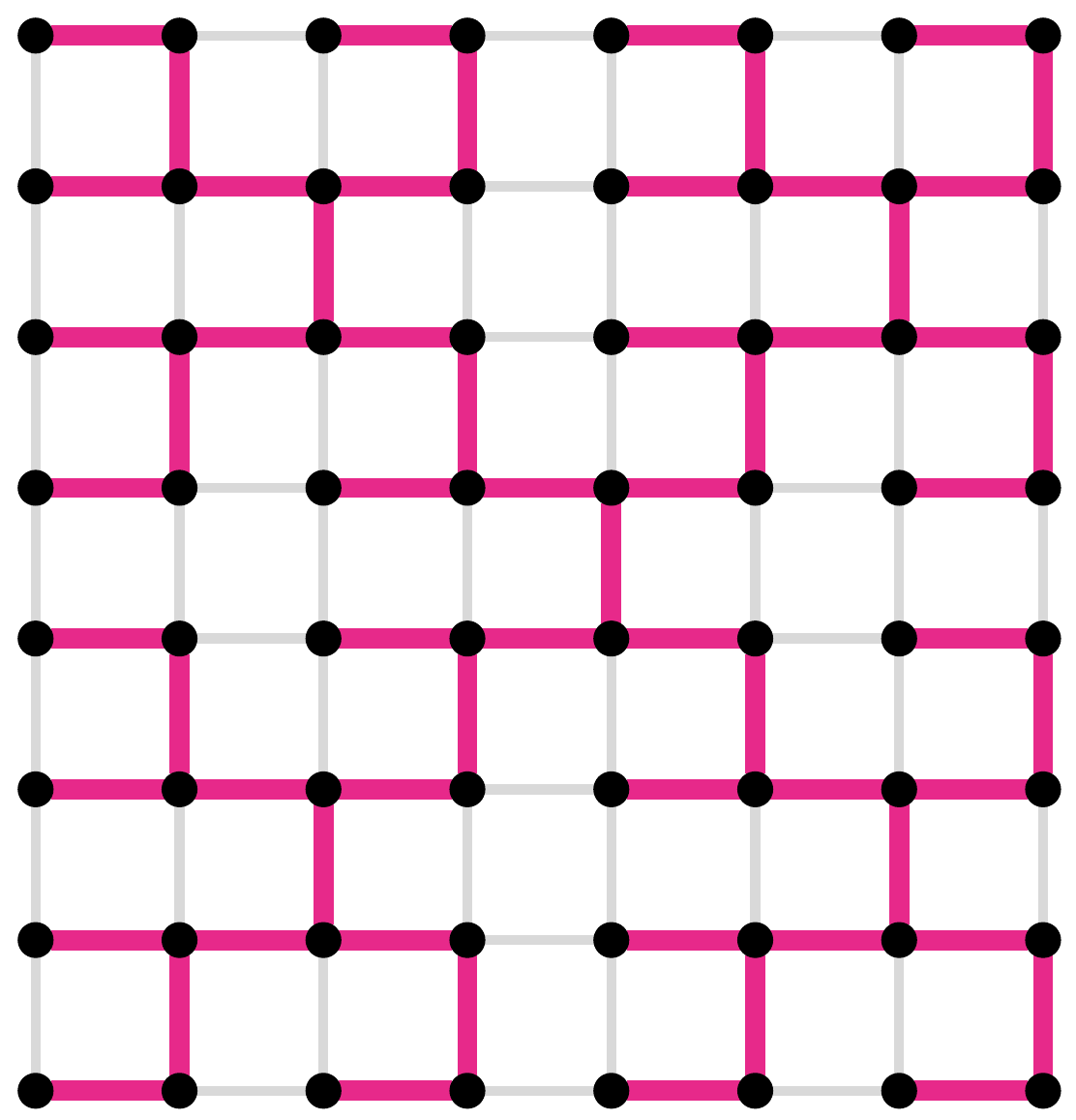}
                \caption{Low-stretch spanning tree, as shown in Alon et al.~\cite{AKPW95}.}
                \label{fig:grid-lst}
        \end{subfigure}
        
        \caption{Comparison between an arbitrary spanning tree and a low-stretch spanning tree for an 8-by-8 grid graph.}
        \label{fig:grid}
\end{figure}

Low-stretch trees are considerably more effective at capturing the structure of general graphs than typical spanning trees. Their power can be illustrated through a simple example in \autoref{fig:grid}. Consider a square grid graph, or mesh, with $n$ vertices. Arbitrary spanning trees generally do a poor job of capturing the structure of this graph, as in the example of the comb shaped tree in Figure 3a. For any vertical edge in the right half of the grid, the endpoints of that edge are at distance 1 in the original graph, but they are at distance $\Omega(\sqrt{n})$ in the tree because the unique path through the tree connecting those two vertices traverses all the way to the left-most column. It is by no means obvious whether any subtree of the grid graph can have substantially better stretch. Alon et al. \cite{AKPW95} pointed out that the grid contains a fractal-like subtree with stretch $O(\log n)$, shown in \autoref{fig:grid-lst}, that is reminiscent of Hilbert's space-filling curve \cite{H91}. They also show that the logarithm is necessary: every spanning tree of an $n$-vertex square grid has stretch $\Omega( \log n )$. Therefore, low-stretch trees provide an opportunity to vastly improve utility of quasi-tree based methods due to provable guarantees of minimizing structural discrepancies.

\subsection{Extracting Low-Stretch Trees From a Graph}
In order to extract a low-stretch tree from an existing graph, we can simply use the existing algorithm from Alon et al.~\cite{AKPW95}. This algorithm performs an iterative coarsening process in order to compute the low-stretch tree. At each step of the algorithm, vertices are partitioned into clusters. Each of these clusters has low topological diameter. Next, a shortest-paths spanning tree is computed for each cluster. The edges from these trees are added to an (initially empty) low-stretch tree, and each cluster is then contracted into a meta-vertex. Edges are then added to represent connections between vertices in different clusters. The algorithm then iterates on this new multigraph.

\subsection{Complexity Analysis}
\label{lst-complexity}

The low-stretch tree construction algorithm from Alon et al.~\cite{AKPW95} runs in time $O(m \log n)$, where $m$=$|E|$ and $n$=$|V|$. The tree is guaranteed to have low average stretch: $\exp(O(\sqrt{\log n\log\log n})) = O(n^{0.01})$. Note that $n^{0.01}$ is very small: it is 1.12 for a graph of one million vertices. 
Other methods with better theoretical guarantees are known~\cite{ABN08,EEST05}, but their algorithms are not practical to implement. A new method claims to have faster running time of $O(m \log n \log\log n)$~\cite{AN12}, but it is unclear whether this approach is practical. Regardless, any method can be substituted to compute the low-stretch tree in this step, provided there are guarantees on the stretch of the resulting tree.

\subsection{Assessing Bundle Quality}

%So far, we have demonstrated how low-stretch trees can be used to compute quasi-tree backbones that capture important aspects of graph structure, even for graphs that at first glance do not look very “tree-like”. We furthermore propose to use them for both quasi-tree layout and for optimal quasi-tree bundling. \tm{do two previous sentences belong elsewhere?? let's discuss!} 
Computing a low-stretch quasi-tree backbone addresses two requirements for edge bundling proposed in previous work~\cite{BS15, PNK11}:  short paths are desirable when bundling edges and sparsity is important for spanners used to route edge bundles. As we explain in Section~\ref{sec:lst}, low-stretch trees offer provable path-shortness guarantees, addressing the first criterion, and our LSQT algorithmic approach is provably sparse compared to arbitrary spanners, addressing the second one.

\section{Related Work}

Our algorithm combines %ideas from several prior methods and contributes to 
the previously distinct spaces of graph layout and bundling approaches, and introduces low-stretch trees to the visualization literature.  
%We also give an overview of other important work on edge bundling. In addition, we discuss previous work on several visualization techniques that we use to display the LSQT algorithm in our proof-of-concept viewer. 

\subsection{Layout}

Quasi-tree layouts have received only limited attention in the previous work; they have not even been considered a major category by previous graph visualization surveys~\cite{lok2001survey, pavlopoulos2008survey}. An early example is the interactive H3 system~\cite{munzner1997h3} that extracts a spanning tree from a general graph and uses it for layout. The backbone edges are drawn at all times using fully custom tree layout that exploits the mathematical properties of hyperbolic geometry. The remainder edges can be toggled on or off for individual nodes, all nodes within a subtree, or the entire graph. Boutin et al.~\cite{boutin2006focus} use the combination of filtering and clustering to extract a ``tree-like graph'' and use that quasi-tree either to guide a standard force-directed layout algorithm or as input for their customized multi-level silhouette tree layout. The LGL system uses a spanning tree as a skeleton to guide their variant of force-directed layout for general graphs~\cite{adai2004lgl}. The multi-stage static SPF layout algorithm~\cite{archambault2006smashing} uses an input spanning tree as an initial skeleton in an extended version of LGL at one of its stages. 

Bourqui and Auber~\cite{bourqui2009large} use a more sophisticated clustering-based approach to extracting quasi-tree structure from a graph in order to draw large quasi-trees at higher quality than SPF. Giot and Bourqui~\cite{giot2015fast} introduce more efficient algorithms for both extracting appropriate quasi-tree structure and a custom bottom-up area-aware layout. 
%\me{this will be impacted by SBEB paper. TM: addressed} 
These two algorithms are similar to our own work since they both include a final bundling stage to reduce clutter, based on Holten's HEB approach~\cite{H06}. Although they do combine layout and bundling for quasi-trees, 
%they do not explicitly compute bundles as first-class data abstractions and 
their fundamental emphasis is on bundling to improve the layout of quasi-trees in specific, whereas ours is to use quasi-trees to drive a bundling algorithm for general graphs. 

\subsection{Bundling}
%In this section we detail existing edge-bundling approaches and methods. We categorize previous approaches into three major families: image-based, geometry-based, and cost-based bundling. 
%We discuss the disparate methods of evaluation used by the authors of each approach as context for our own evaluation choices in this paper.

Edge bundling approaches fall into three major families: image-based, geometry-based, and cost-based bundling.
Image-based bundling \cite{ersoy2011skeleton, zhou2013edge} uses processes like splatting and shape skeletonization on previously computed bundling layouts (from either cost-based or geometry-based approaches) to build new shaded shapes used for visualizing each set of edges. 
%Image-based approaches
%, which handle both directed and undirected graphs, 
%typically leverage GPU acceleration to produce highly organic-looking, smooth bundles, which follow a tree structure that facilitates the ability to see how edges join, split, and flow into major structures. These approaches can be advantageous in highlighting local hierarchies in data, counting and comparing branch size, and distinguishing bundles. 
%Some of the advantages of image-based approaches are that they intuitively leverage existing image rendering tools and techniques, and that they can be used together with other bundling layout approaches ~\cite{zhou2013edge}. 
%Beyond these implementation advantages, efficiency, and aesthetics, there is no obvious task-based reason to choose Image-based approaches over any other edge bundling style. 
Geometry-based bundling \cite{cui2008geometry, qu2006controllable} employs spatial decomposition using spatial data structures such as quadtrees, uniform or non-uniform grids, or triangle meshes to determine the shape and curve of the edges in the display. 
%Since not every graph is comprised of a meaningful hierarchy, grids are often used to guide the bundling. Additional structures include triangle meshes, non-uniform grids, and quadtrees \cite{cui2008geometry}. 
%For example, Qu et al.\cite{qu2006controllable} generate a control mesh using Delaunay triangulation, which takes vertices (nodes from the original graph) as input. In this case, instances of Delaunay edges intersecting with graph edges become control points for the edge bundling. The control mesh can be adjusted in order to facilitate interactive control and clustering over the edges.
Cost-based bundling \cite{HET12, gansner2011multilevel} is named based on the metaphor of saving the cost of ink, and more generally includes all energy minimization approaches. LSQT falls into this category. The shapes of the edges are determined by the cost of ink, or energy, it will take to draw the edges. Multilevel agglomerative edge bundling (MINGLE) \cite{GHNS11} is one example of an ink-saving strategy. Force-directed edge bundling methods~\cite{holten2009force} are all examples of energy minimization approaches, which focus on reducing the spring energy associated with the models used to draw a given system \cite{GHNS11}. Cost-based approaches typically aim for aggressive de-cluttering over maintaining the interpretability of the existing spatial topology.

Three cost-based approaches are most similar to our own. One is the seminal edge bundling proposal, Hierarchical Edge Bundling (HEB)~\cite{H06}, which does not rely on an existing layout; we call this property \textbf{layout agnostic}. Surprisingly, despite the enormous amount of followup work on edge bundling, no later methods extended this aspect of the work. Instead, the many examples 
\cite{HET12, hurter2012graph, ersoy2011skeleton, qu2006controllable, gansner2011multilevel, lambert2010winding, BS15, PNK11} of subsequent methods for bundling are in the category we  
call \textbf{layout first}, where they rely on an existing layout in order to compute bundles. To the best of our knowledge we are the first to continue exploring the benefits of layout-agnostic approach, which avoids the computational expense of a preprocessing step to create an initial graph layout and allows users to explore multiple possible layout approaches for the underlying quasi-tree structure without the need to recompute bundles. 

Kernel Density Estimation Edge Bundling (KDEEB) \cite{HET12} is a cost-based layout-first approach that improved upon several documented computational complexity issues with other algorithms for general graphs by providing a robust and simple approach, with an implementation that performs faster than most other algorithms. Edges are advected along a density map to create a layout with smooth and well-separated bundles. The visual results are notably different than many other edge bundling approaches, so direct comparison to previous methods is difficult. The authors argue that a useful quality heuristic is to have areas of clear separation between high-density bundles where the edges are densely packed, given the goal of minimizing ink~\cite{HET12}. Our LSQT approach takes a similarly aggressive approach to de-cluttering, and also yields results quite visually different from previous methods. 

Clustered Edge Routing (CER) \cite{BS15} is also a cost-based, layout-first approach. It is the most similar approach to LSQT in terms of using a sophisticated mathematical formalism to compute spanners, well-separated pair decomposition, in order to achieve aggressive de-cluttering of graph visualizations. The authors quantify several desirable properties of a routing graph, and two of them can be applied to our layout-agnostic approach: sparsity is crucial for routing graphs, and the shortest path between two vertices in the routing graph should not be much longer than their direct connection in the graph. In our work, we use these properties as guidelines for the characterization of desirable properties of spanners for bundle routing. Specifically, our low-stretch spanning tree has provable guarantees of sparsity that are stronger than the limited guarantees provided by their method of computing spanners: 
to preserve distance with a multiplicative factor $O(t)$, it is known that $\Omega(n^{1+1/t})$ edges are needed~\cite{ADDJ90}. Also, the complexity of their method is $O(n^2 \log n)$; we have clear speed advantages, in that both the computational complexity and the empirical performance of our algorithm are superior.

\subsection{Low-Stretch Trees}

The mathematical formalism of low-stretch trees~\cite{AKPW95} arose in the theoretical computer science community where it has been applied to the construction of spanners, but has not been previously introduced in the visualization or graph drawing literature. 

\section{Quasi-Tree Bundling Algorithm}

\begin{figure*}[t]
%  \vspace{-15pt}
    \centering
    \begin{subfigure}[t]{0.3\textwidth}
        \centering \includegraphics[width=.6\textwidth]{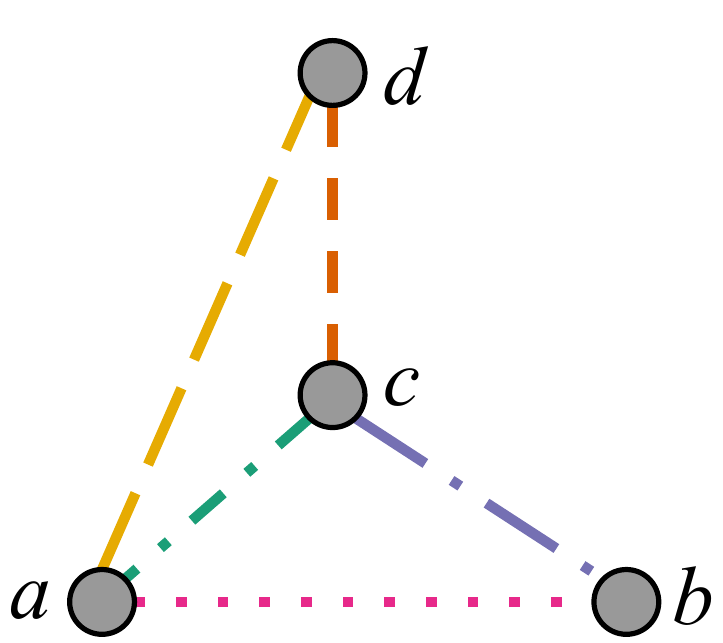}
        \caption{}
        \label{fig:bundle:a}
    \end{subfigure}
    ~
    \begin{subfigure}[t]{0.3\textwidth}
        \centering \includegraphics[width=.6\textwidth]{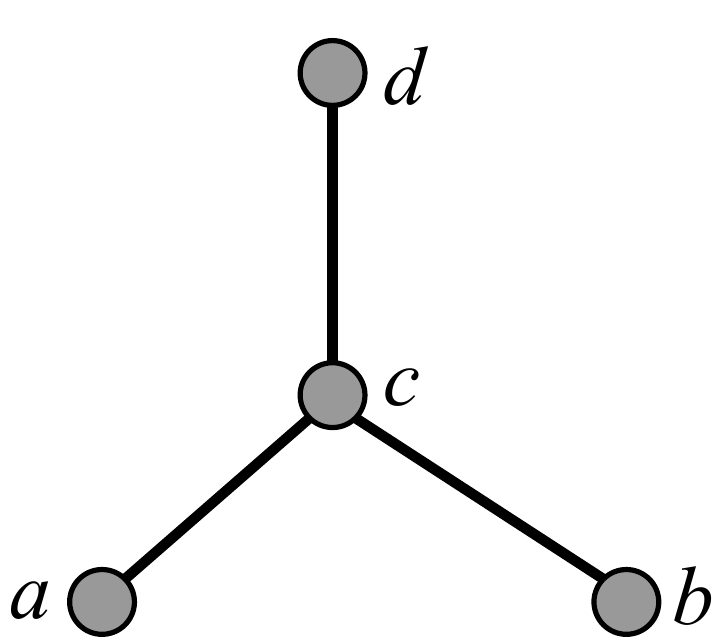}
        \caption{}
        \label{fig:bundle:b}
    \end{subfigure}
    ~
    \begin{subfigure}[t]{0.3\textwidth}
        \centering \includegraphics[width=.6\textwidth]{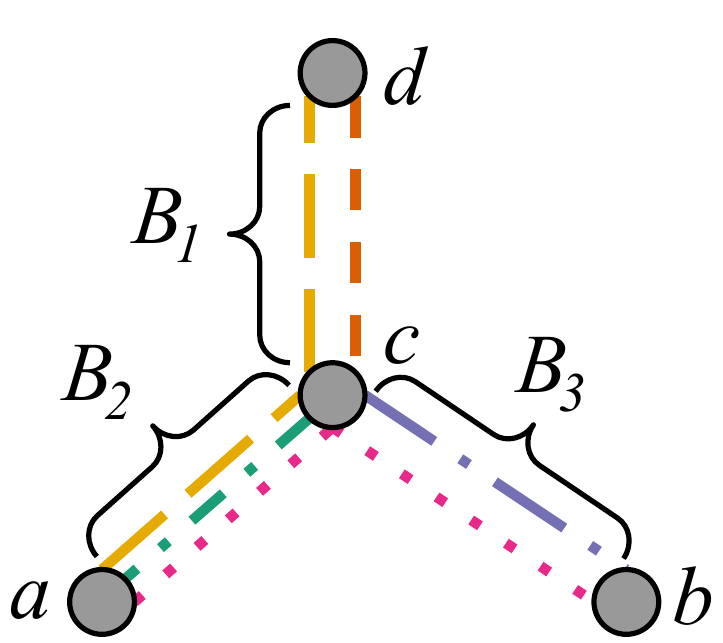}
        \caption{}
        \label{fig:bundle:c}
    \end{subfigure}
        \caption{LSQT bundling algorithm. A low-stretch tree $T=(V,E_T)$ is computed for use as a routing graph. Segmentation is performed by routing edges through this tree, as illustrated in \hyperref[fig:routing]{Figure \ref*{fig:routing}}. Bundles are then formed from groups of segments sharing the same endpoints as edges in $E_T$. (a) Original graph $G$, where colour and dash style are used to differentiate edges. (b) Routing tree $T$. (c) Result showing three bundles, where segments have the same visual encoding as above.}
    \label{fig:bundle}
\end{figure*}

\begin{figure*}[h!]
    \centering
       \begin{subfigure}[t]{0.3\textwidth}
    \centering
        \includegraphics[width=.6\textwidth]{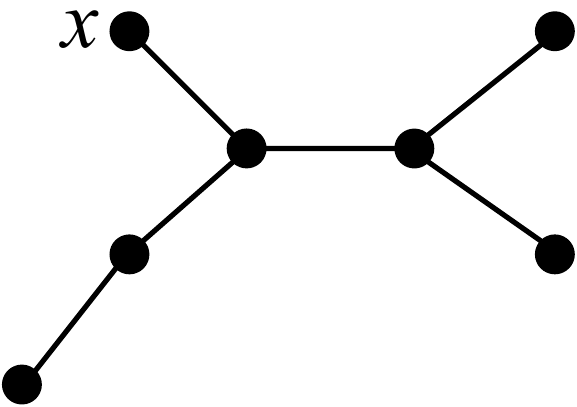}
        \caption{}
        \label{fig:routing:a}
    \end{subfigure}
    ~
    \begin{subfigure}[t]{0.3\textwidth}
    \centering
        \includegraphics[width=.6\textwidth]{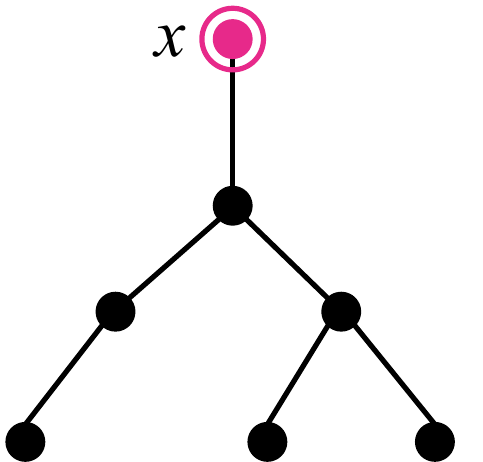}
        \caption{}
        \label{fig:routing:b}
    \end{subfigure}
    ~
    \begin{subfigure}[t]{0.3\textwidth}
    \centering
        \includegraphics[width=.6\textwidth]{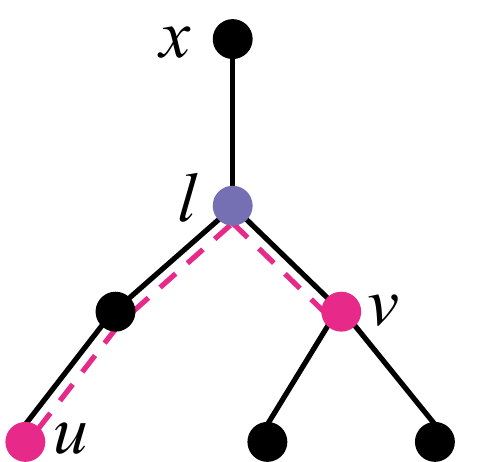}
        \caption{}
        \label{fig:routing:c}
    \end{subfigure}
    \caption{LSQT routing algorithm. The tree $T = (V,E_T)$ is rooted at an arbitrary vertex in order to speed up queries, which then only need to find the lowest common ancestor between two vertices. (a) Original (unrooted) tree $T$. (b) $T$ rooted at arbitrary vertex $x$. (c) 
    To find the path from $u$ to $v$ in $T$, we find their lowest common ancestor, $\ell$, by climbing up the tree. The path from $u$ to $v$ is then the concatenation of paths $u\text{-}\ell$ and $\ell\text{-}v$.}
    \label{fig:routing} 
\end{figure*}

Our method computes bundles directly, without relying on the geometry of vertices or edges. We route edges through a tree to determine both edge segmentation and bundle membership. The algorithm consists of two steps: routing graph generation, where the spanning tree is computed; and edge routing, where edges are segmented and bundles are formed. We first detail how to segment remainder edges, and then explain how our process allows for subsequently efficient path queries in the routing process. Finally, we present a complexity analysis for LSQT. 

\subsection{Efficient Segmentation}
We will use a low-stretch tree as our routing graph for bundling. We compute this tree using the algorithm proposed by Alon et al.~\cite{AKPW95}. It performs an iterative coarsening process in order to compute the low-stretch tree. We subdivide each remainder edge into segments that follow the backbone quasi-tree vertex structure. We bundle only these remainder edges, and each remainder edge is represented by an ordered list of segments.  This process is illustrated in \autoref{fig:bundle}.

At each step of the algorithm, the vertices of the graph are partitioned into clusters such that each cluster has low topological diameter. A shortest-paths spanning tree is then computed for each cluster, and the edges from these trees are added to the (initially empty) low-stretch tree. Each cluster is then contracted into a meta-vertex, and edges are added to represent connections between vertices in different clusters. The result is a multigraph, where a single node may share multiple edges. This multigraph contains only quasi-tree edges, where the nodes are vertices from the original graph. The multigraph is used to create bundles, which are simply defined as a set of two or more segments with the same endpoints. LSQT then iterates on this new multigraph. Segments in the same bundle are referred to as bundle neighbors. The construction of these segments is thus purely topological and does not depend on geometric layout. 

To perform segmentation and bundling efficiently, we query $u\text{-}v$ paths in our spanning tree $T$. We need a fast subroutine to handle this procedure. A na{\"i}ve approach is to simply perform breadth-first search, but that would require $\Theta(|V|)$ time per query. Since $\Omega(|E_R|)$ queries are needed, where $E_R$ are the remainder edges, this approach requires $\Omega(|V|\cdot|E_R|)$ time. Instead, we will use a two-phase approach that pre-processes the graph in linear time, but can then perform queries in time proportional to the length of the returned path (which is optimal, as the path itself is returned). Performing such a query for every remainder edge in $E_R$ requires time $O(s_T(G)\cdot|E_R|)$, which is small since $T$ is a low-stretch tree.

This subroutine can be easily implemented as follows. The pre-processing step simply roots the tree at an arbitrary vertex and directs all edges towards the root. Then, to find a $u\text{-}v$ path in $T$, step \emph{in parallel} from each of $u$ and $v$ towards the root, marking the nodes along the way. The first marked vertex encountered on either path is the lowest common ancestor $\ell$, and the time to identify it is proportional to the length of the $u$-$v$ path.
The path through the tree is then the concatenation of paths $u\text{-}\ell$ and $\ell\text{-}v$, and the segmentation of $(u,v)$ is this path. This process is illustrated in \autoref{fig:routing}. Once the $u\text{-}v$ paths have been computed for each $(u,v)$, bundles are formed such that all segments that have the same endpoints are bundle neighbours.

\subsection{Complexity Analysis}
The query time to identify the least common ancestor (LCA) is proportional to path length. As stated in Section~\ref{lst-complexity}, the time to compute the low-stretch tree is $O(m \log n)$ where $m$=$|E|$ and $n$=$|V|$. As observed above, the total time for querying the low-stretch paths for every remainder edge in $E_R$ requires time $O(s_T(G)\cdot|E_R|) = O( m * n^{0.01} )$.

\section{Visual Encoding and Interaction Possibilities}
In this section, we discuss a proof-of-concept viewer to demonstrate visual encoding possibilities  enabled by the LSQT algorithm. We take advantage of the key assets of our bundling algorithm when designing the visual encodings for LSQT: namely, that it does not require a pre-determined layout, and that it creates the data abstraction of bundles that are directly accessible through their own data structure. The viewer supports interactive queries of the bundle structure, and also allows arbitrary layout adjustment through dragging. 

\subsection{Quasi-Tree Layout Approaches}
\begin{figure}[h!]
       \centering

  \label{fig:layout}
        \begin{subfigure}[t]{0.45\textwidth}
                \includegraphics[width=\textwidth]{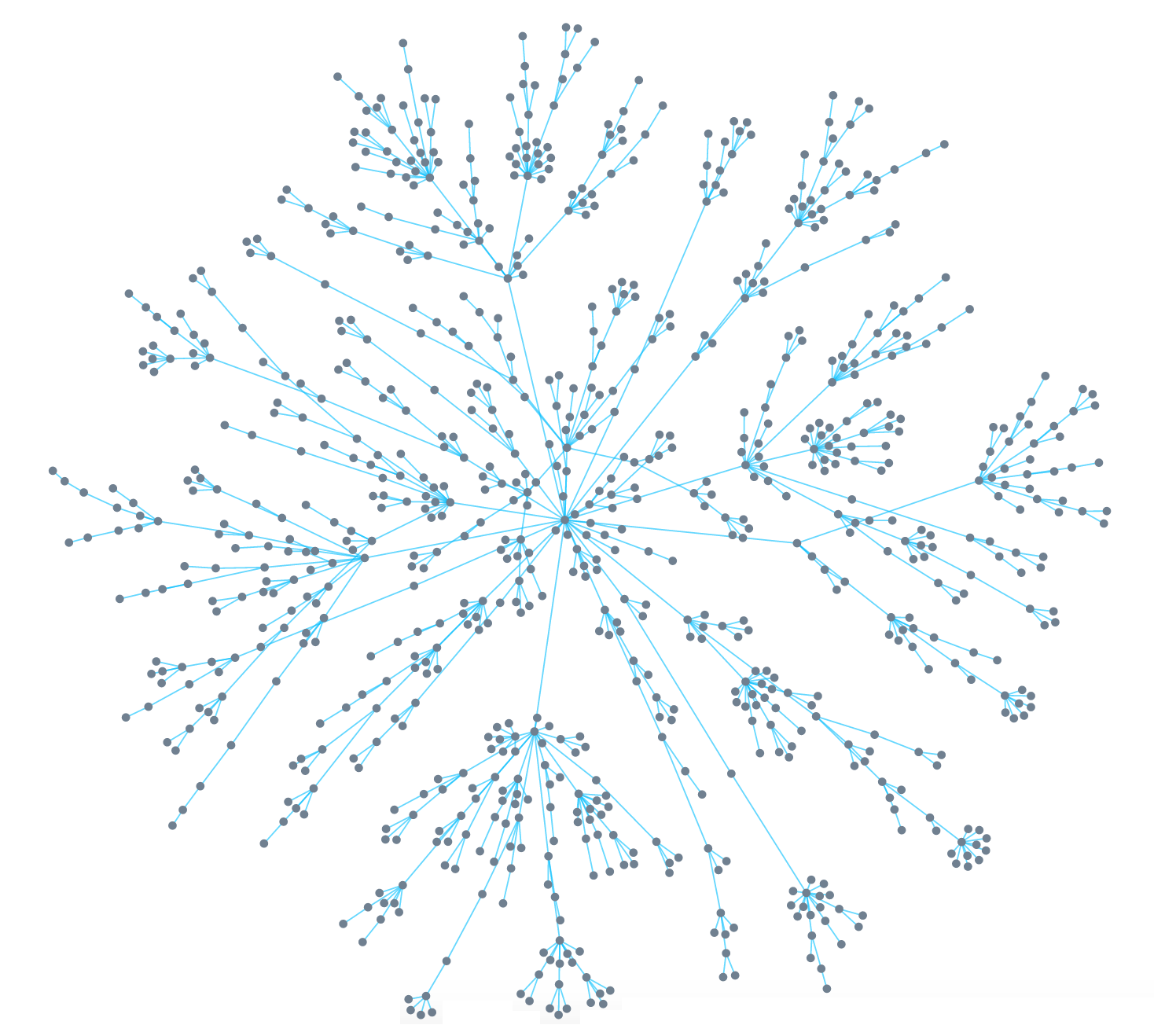}
                \caption{Force-directed layout}
                \label{fig:layout:force}
        \end{subfigure}
        ~ %add desired spacing between images, e. g. ~, \quad, \qquad, \hfill etc.
        \begin{subfigure}[t]{0.45\textwidth}
                \includegraphics[width=\textwidth]{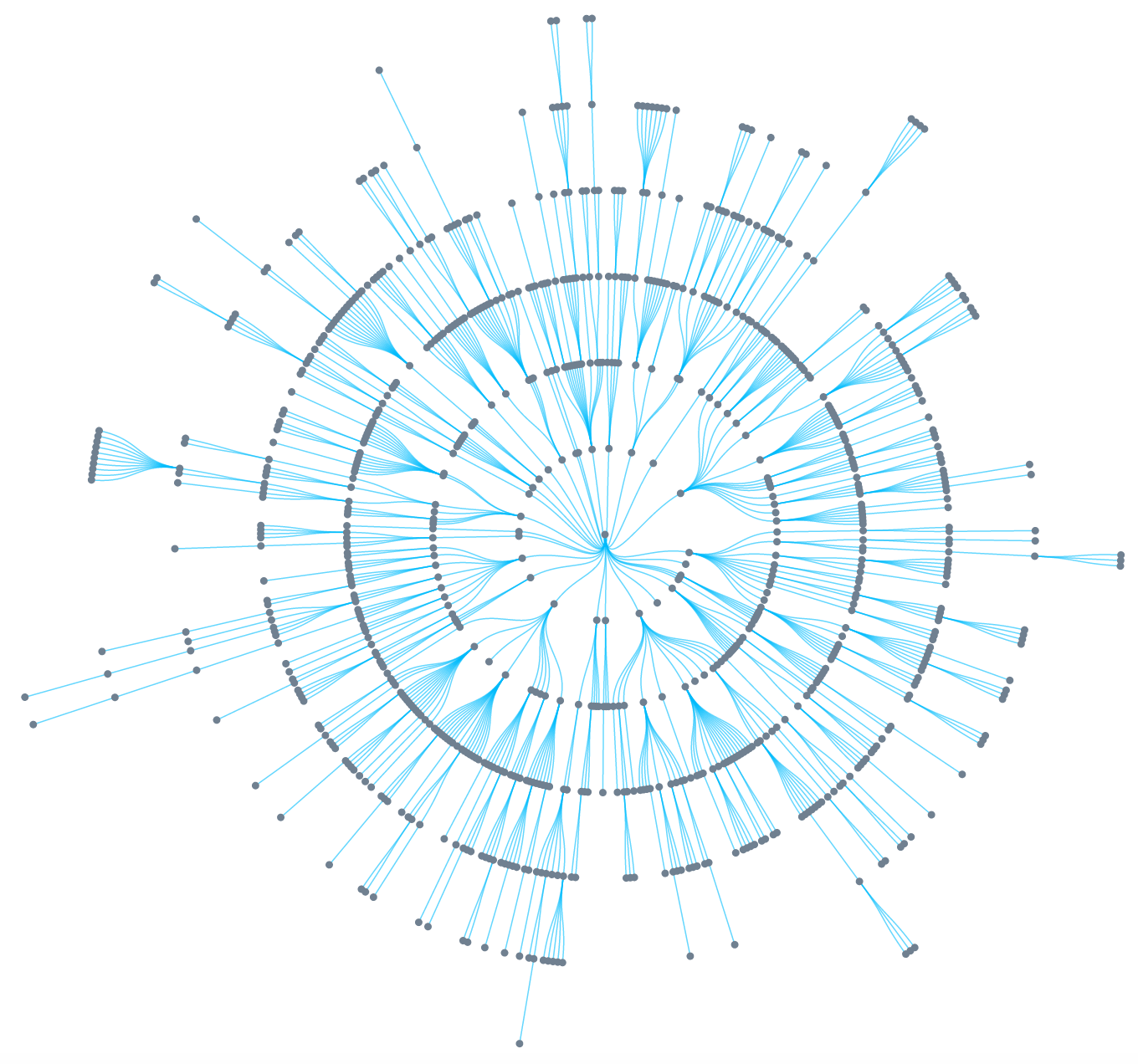}
                \caption{Radial Reingold-Tilford layout}
                \label{fig:layout:tree}
        \end{subfigure}
         \caption{Comparison between two common vertex layout possibilities for the LSQT routing graph $T$.}
         \end{figure}

Because LSQT computes bundles before layout, we can position the vertices in any way we choose. For example, we can use any layout for the low-stretch tree backbone. \autoref{fig:layout:force} and \autoref{fig:layout:tree} show a proof-of-concept viewer using two popular layout approaches: the standard force-directed graph layout that is built into D3~\cite{BOH11}, and the standard Reingold-Tilford tree drawing approach~\cite{RT81}. Future work could use any previously proposed or custom-designed tree drawing approach, since the LSQT backbone is itself a tree. 

\subsection{Encoding and Interaction}

In addition to the layout, there are remaining design choices for visual encoding and interaction. 

\subsubsection{Bundle Visual Encoding}
%Previous edge bundling methods draw edges individually, so bundles are shown as closely located or overlapping edges. However, we wish to also take advantage of the fact that our algorithm outputs bundles, and that all segments in a bundle share the same endpoints. 

\begin{figure}[h]
       \centering
        \begin{subfigure}[t]{0.45\columnwidth}
       \centering
                \includegraphics[width=\textwidth]{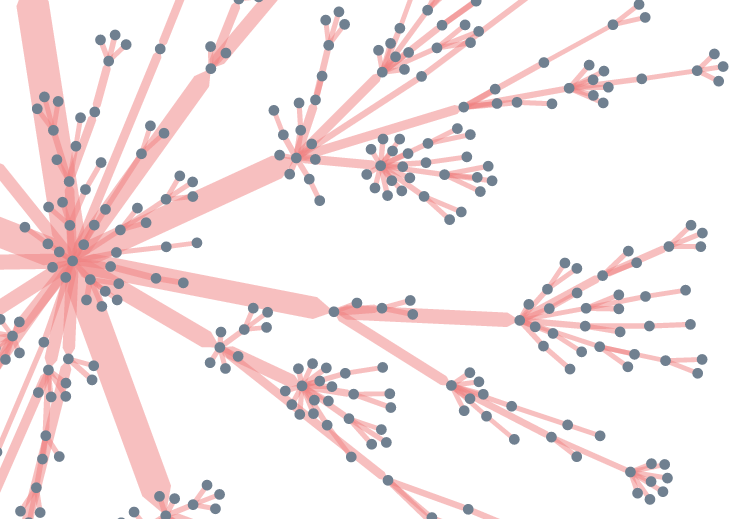}
                \caption{Bundles only}
                \label{fig:vis-enc:bu}
		  \vspace{3pt}
        \end{subfigure}
        ~
        \begin{subfigure}[t]{0.45\columnwidth}
       \centering
                \includegraphics[width=\textwidth]{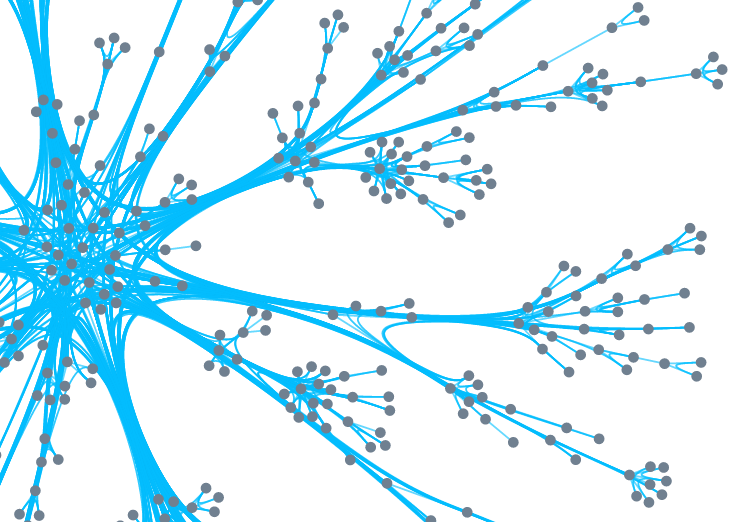}
                \caption{Remainder edges only}
                \label{fig:vis-enc:re}
		  \vspace{3pt}
        \end{subfigure}
        \begin{subfigure}[t]{0.45\columnwidth}
       \centering
                \includegraphics[width=\textwidth]{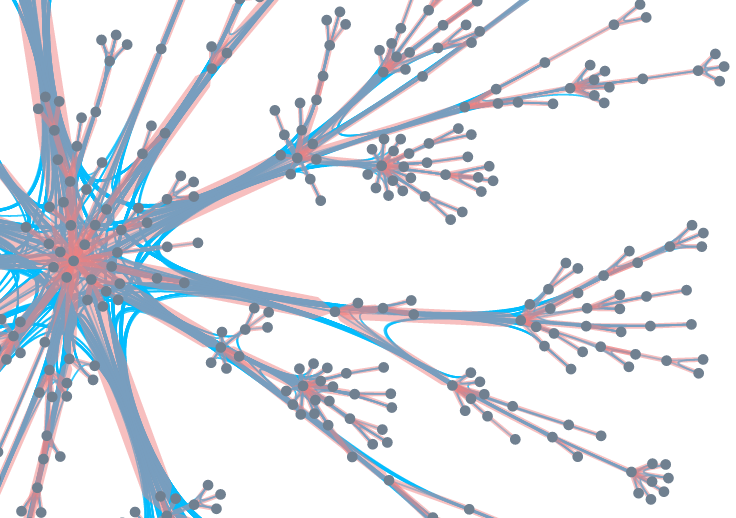}
                \caption{Bundles foreground}
                \label{fig:vis-enc:bu-top}
        \end{subfigure}
        ~
        \begin{subfigure}[t]{0.45\columnwidth}
       \centering
                \includegraphics[width=\textwidth]{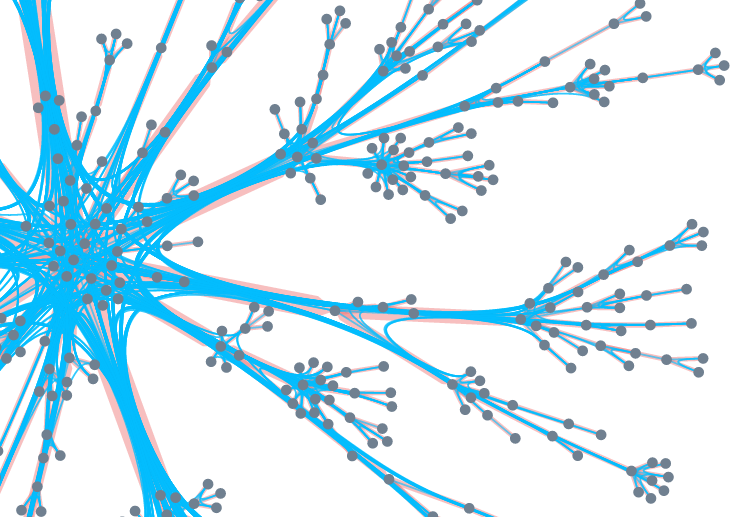}
                \caption{Remainder edges foreground}
                \label{fig:vis-enc:re-top}
        \end{subfigure}
  \vspace{-5pt}
  \caption{Examples of visual encoding options for LSQT.}
  \label{fig:vis-enc}
\end{figure}

We draw bundles explicitly, both independently from and in combination with individual drawing of remainder edges.  \autoref{fig:vis-enc} shows the different visual encoding options for bundles and edges. Bundles are depicted by straight lines with tapered endpoints and varying thickness, as shown in  \autoref{fig:vis-enc:bu}. The thickness of bundles varies in proportion with bundle size (the number of segments in a bundle). Individual edges are drawn as splines, as shown in \autoref{fig:vis-enc:re}. The control points for the spline of edge $(u,v)$ are the endpoints of its segments, which correspond to the vertices in the $u\text{-}v$ path in $T$. This approach is similar to the method of HEB \cite{H06}.
Bundles and edges can also be drawn simultaneously, where distinct perceptual layers are created by adjusting opacity.  \autoref{fig:vis-enc:bu-top} and \autoref{fig:vis-enc:re-top} show the difference between having bundles and remainder edges as the foreground layer.

\subsubsection{Interaction Idiom Examples}
\begin{figure}[h]
       %\centering
        \begin{subfigure}[t]{0.45\columnwidth}
       \centering
                \includegraphics[width=\textwidth]{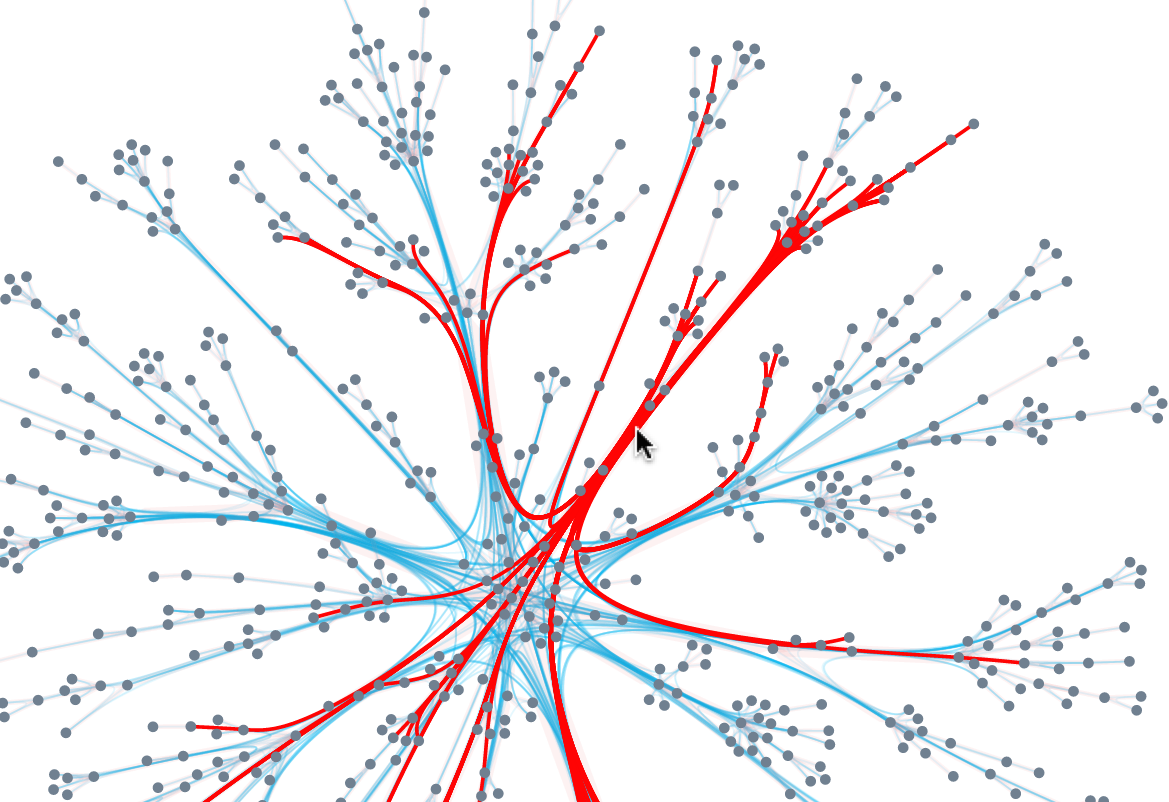}
                \caption{Hovering over a bundle highlights edges in that bundle}
                \label{fig:hover:bu}
        \end{subfigure}
        ~ %add desired spacing between images, e. g. ~, \quad, \qquad, \hfill etc.
        \begin{subfigure}[t]{0.45\columnwidth}
       \centering
                \includegraphics[width=\textwidth]{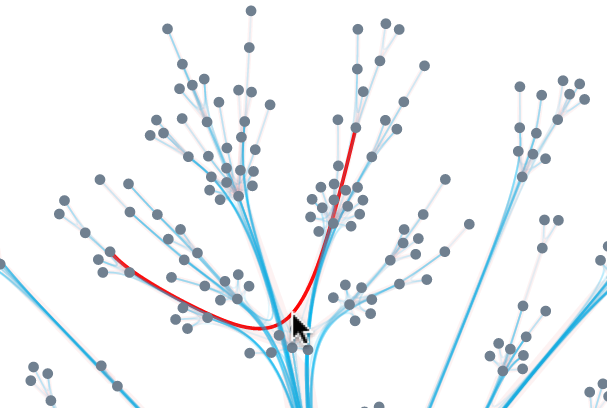}
                \caption{Hovering over an individual edge highlights that edge}
                \label{fig:hover:re}
        \end{subfigure}
  \vspace{-5pt}
  \caption{Comparing bundle querying and edge querying.}
  \label{fig:hover}
\end{figure}

%Our bundling method vastly increases support for user interaction. 
In our proof-of-concept viewer, vertices can be re-positioned by the user with interactive dragging, without the need to re-run the edge bundling algorithm. Previous bundling methods would not support this style of interaction because their bundling is dependent on a static and pre-computed layout. 
%Different hovering techniques are also possible with our explicitly computed bundles that are "first-class citizens", where edge membership in bundles is known. 
The viewer also supports interactive exploration of the mapping between edges and bundles. \autoref{fig:hover:bu} shows hovering over a bundle in order to highlight all edges belonging to that bundle in the viewer. Figure \autoref{fig:hover:re} shows  single edges individually highlighted on hover.

%Rewrite above?:
%Bundle vs. remainder edges visually encoded with color
%Bundle thickness varies in proportion with bundle size
%More segments = thicker bundle. 
%Individual edges are drawn as splines
%similar to how HEB handles individual edges
%Control points for a spline of edge (u,v) are the endpoints of it segments (i.e., vertices in the u-v path in the tree).

\subsubsection{Visual Layering}
%Explicit computation of bundles allows users to make interactive decisions about what to draw and view. 
Users can choose to display one of three modes: bundles alone, remainder edges alone, or a combination of bundles and remainder edges simultaneously.
Distinct perceptual layers are created by adjusting the opacity when drawing bundles/backbones and remainder edges simultaneously. Additionally, users can switch between bundles and remainder edges as the foreground layer.

\section{Results}
We present results using datasets that have been used in previous work; we chose them to enable comparisons between existing methods, based on their popularity in cited literature and diversity among layout and bundling style. 
%Our results do not make explicit qualitative comparisons or claims about use cases. We simply wished to demonstrate the visual difference between our method and others, but acknowledge that future work should investigate specific use cases for quasi-tree layout and bundling methods. 
\autoref{tab:graphs} lists the five graphs, along with size statistics (number of vertices and edges) and a brief description.

\setlength{\tabcolsep}{5pt}
\begin{table}[htb]
\begin{center}
    \begin{tabular}{ | l | l | l | l | }
    \hline
    \bf dataset & \bf $|V|$ & \bf $|E|$ & \bf Description \\ \hline
    Flare & 220 & 708 & Software hierarchy\footnotemark[1]\\ \hline
    Poker & 859 & 2127 & Poker game graph\footnotemark[3]\\ \hline
    Email & 1133 & 5451 & Email interchange\footnotemark[4]\\  \hline
    Yeast & 2224 & 6609 & Protein interaction\footnotemark[4] \\ \hline
    Wiki & 7066 & 100736 & Wikipedia elections\footnotemark[4] \\ \hline
    \end{tabular}
\end{center}
  \vspace{-10pt}
\caption{Graph statistics for datasets used in this paper.}
\label{tab:graphs}
\end{table}

\begin{table}[h]
\centering
\begin{center}
    \begin{tabular}{ | l | l | l | l | l | l | l |}
    \hline
    \bf dataset & \bf $|V|$ & \bf $|E|$ & \bf LSQT & \bf Bundle & \bf Draw & \bf Total \\ \hline
Flare & 220 & 708 & 0.022 & 0.005 & 0.032 & 0.060\\ \hline
Poker & 859 & 2127 & 0.082 & 0.026 & 0.108 & 0.216\\ \hline
Email & 1133 & 5451 & 0.180 & 0.053 & 0.283 & 0.516\\ \hline
Yeast & 2224 & 6609 & 0.247 & 0.070 & 0.342 & 0.659\\ \hline
Wiki & 7066 & 100736 & 4.275 & 1.010 & 2.782 & 8.067\\ \hline
    \end{tabular}
\caption{Performance statistics of LSQT. Running time is in seconds, and is averaged over 100 runs.}
\label{tab:performance}
\end{center}
\end{table}

\subsection{Implementation}
LSQT is implemented in Python and JavaScript, using D3~\cite{BOH11} for drawing. Our code is open-source and available at \url{https://github.com/rebvan/lsqt}. Our proof of concept viewer can be found at \url{https://lsqt-vis.herokuapp.com}. All results in this paper are from runs on a MacBook Pro with a 2.6 GHz Intel Core i7 and 16 GB of 1600 MHz DDR3 RAM.

\subsection{Computational Performance}

\autoref{tab:performance} shows performance statistics of LSQT, breaking down the total computation time into three phases: computing the low-stretch backbone, computing the bundles and their data structure, and drawing the graph using the standard D3 force-directed drawing. Running times for LSQT range from 0.06 seconds for a graph with over 700 edges to just over 8 seconds for graph of over 10,000 edges. On this large graph, the MINGLE paper reports a running time of 18.4 seconds, albeit on an older architecture~\cite{GHNS11}; it is the fastest previous method, and we achieve competitive performance with our proof-of-concept implementation that uses an unoptimized scripting languages. 
In contrast, MINGLE is implemented in C using OpenGL on a GPU, leading us to believe that LSQT speed could be substantially improved with a GPU port or simple optimization techniques.

\subsection{Qualitative Layout Comparison}
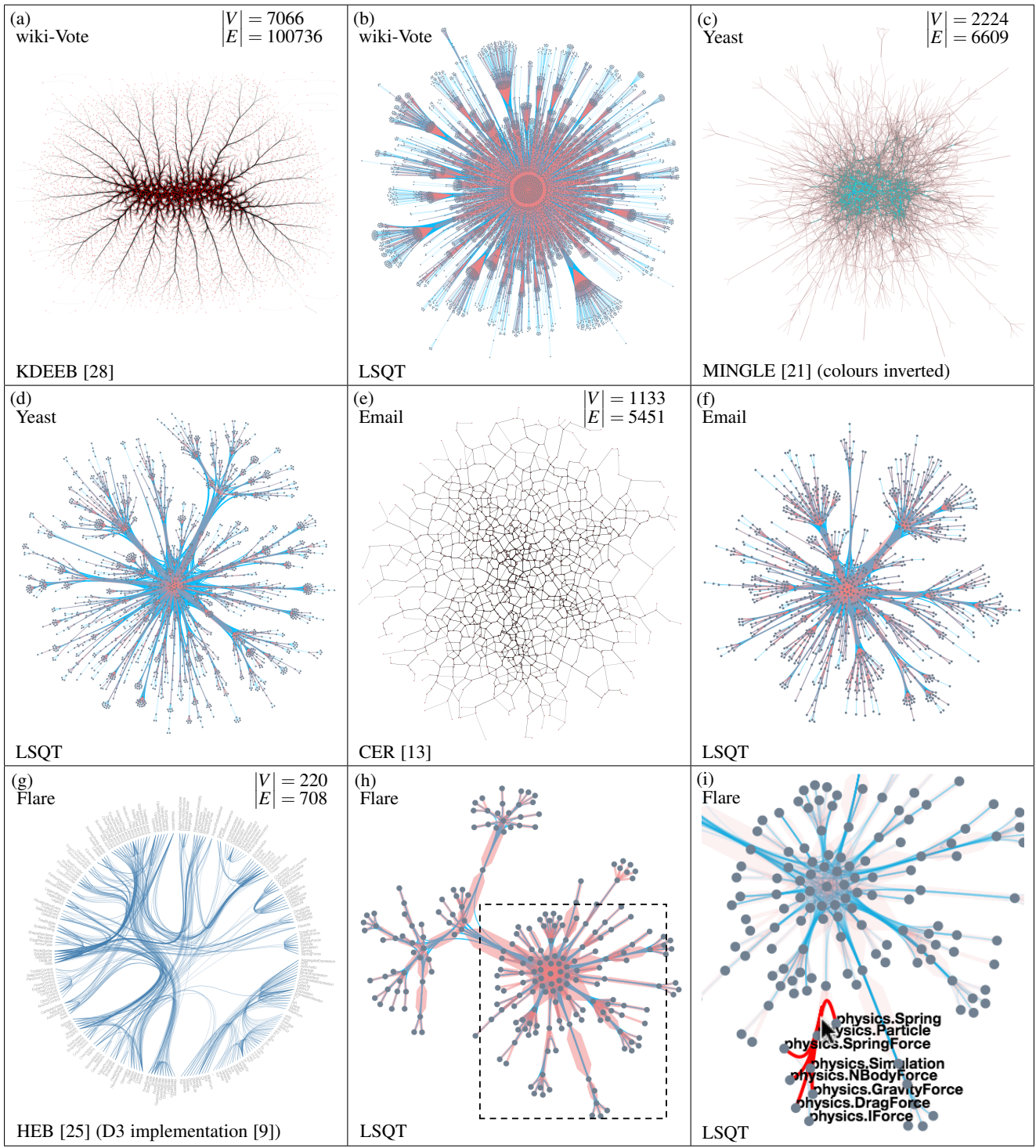
\begin{figure*}[hbtp]
\centering
\begin{tabular}{| C{0.3\textwidth} | C{0.3\textwidth} | C{0.3\textwidth} |}
\hline
\begin{overpic}[width=0.3\textwidth,trim=0 0 0 -5]{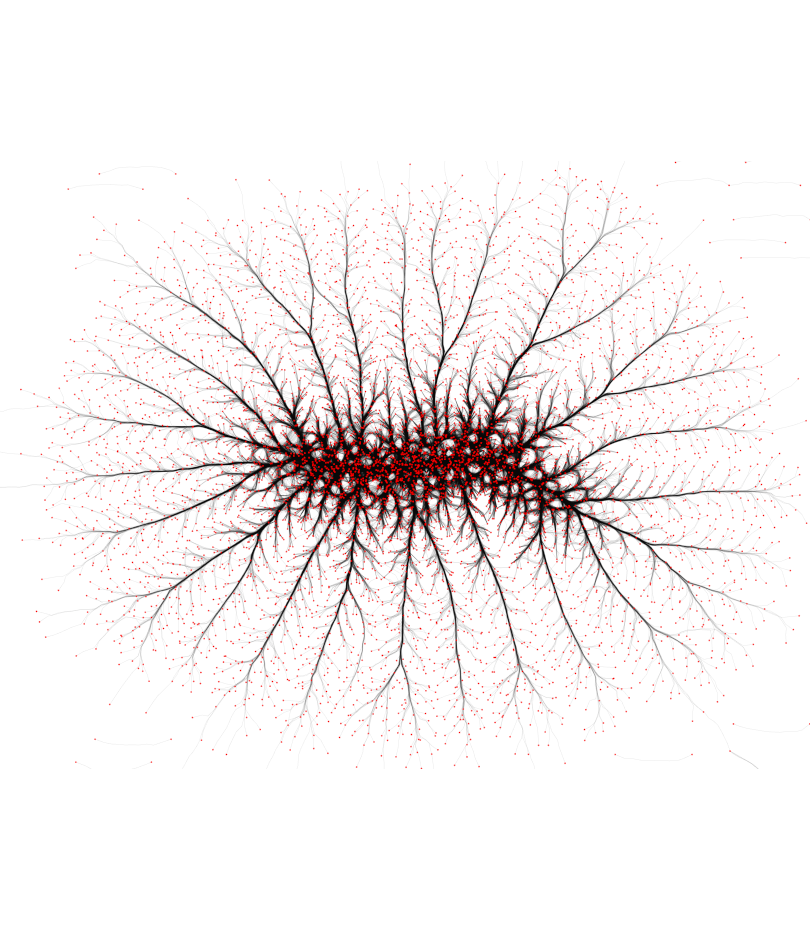}
 \put (-3,95) {\refbigfig{fig:comp:wiki-kdeeb} \small (\alph{bigfig})}
 \put (0,90) {\small wiki-Vote}
 \put (55,95) {\small $|V| = 7066$}
 \put (55,90) {\small $|E|=100736$}
 \put (0,0) {\small KDEEB~\cite{HET12}}
\end{overpic} &
\begin{overpic}[width=0.3\textwidth,trim=0 0 0 -5]{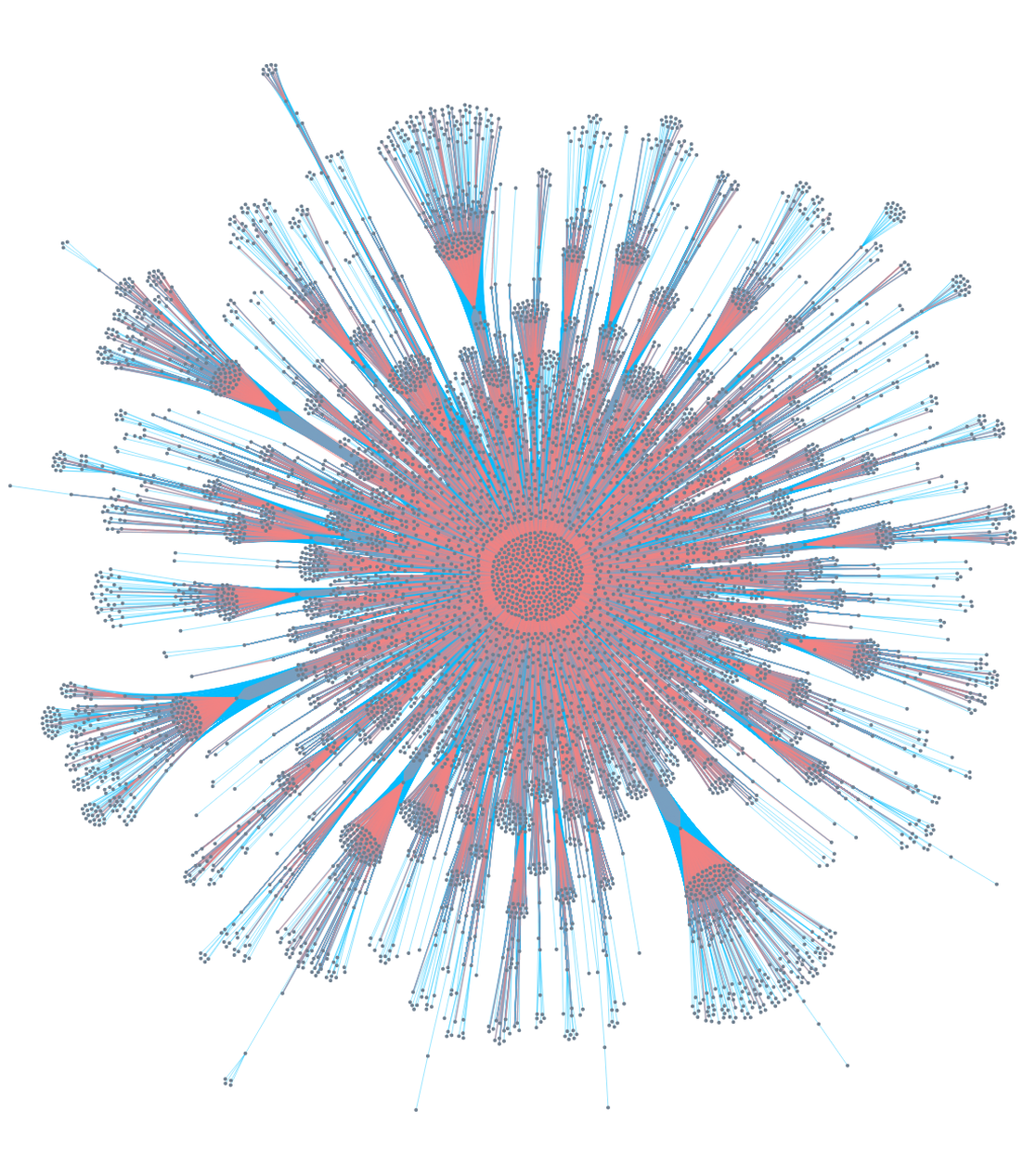}
 \put (-3,95) {\refbigfig{fig:comp:wiki-lstb} \small (\alph{bigfig})}
 \put (0,90) {\small wiki-Vote}
 \put (0,0) {\small LSQT}
\end{overpic} &
\begin{overpic}[width=0.3\textwidth,trim=0 0 0 -5]{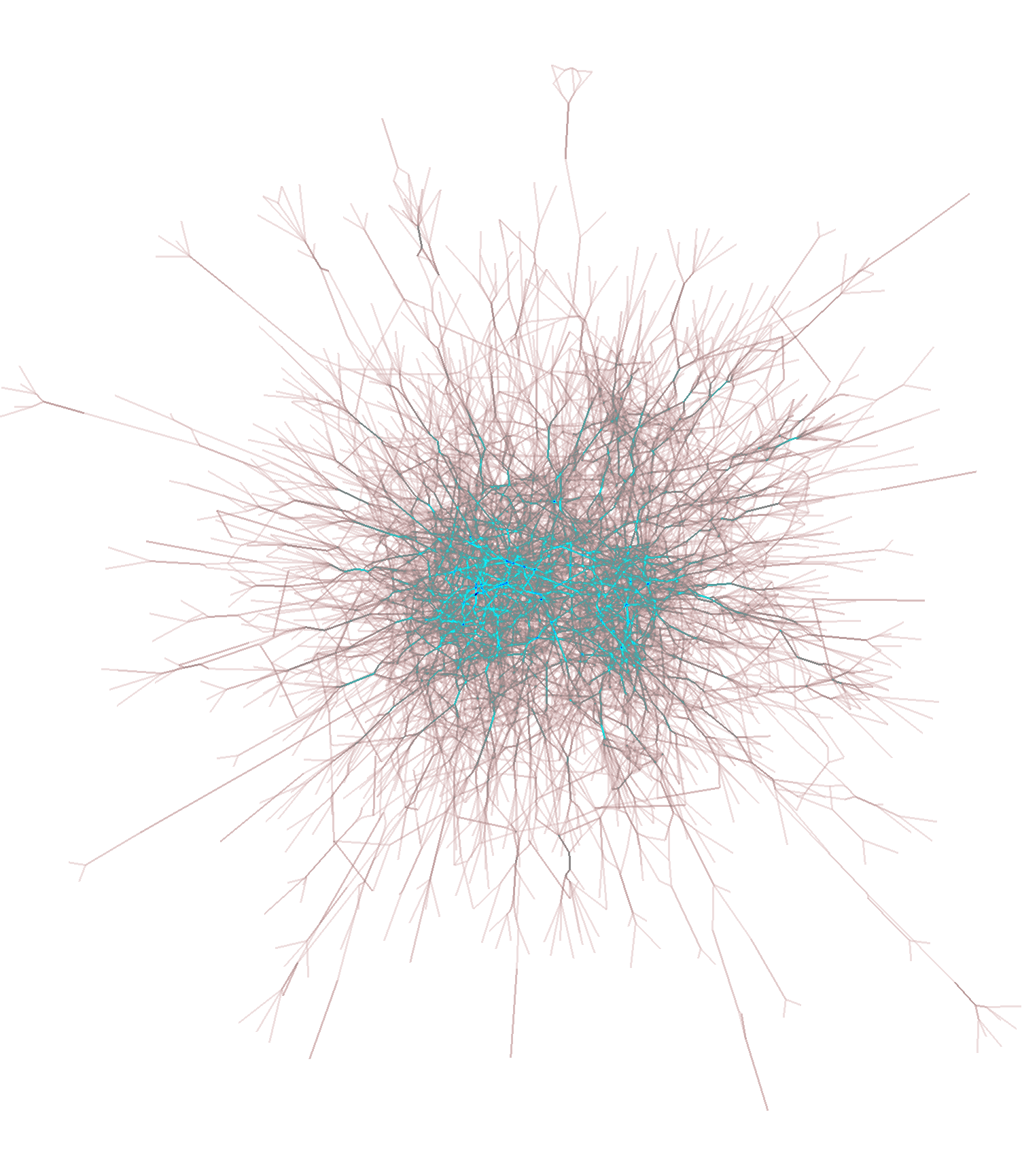}
 \put (-3,95) {\refbigfig{fig:comp:yeast-mingle} \small (\alph{bigfig})}
 \put (0,90) {\small Yeast}
 \put (60,95) {\small $|V| = 2224$}
 \put (60,90) {\small $|E|=6609$} 
 \put (0,0) {\small MINGLE~\cite{GHNS11} (colours inverted)}
\end{overpic} \\
\hline
\begin{overpic}[width=0.3\textwidth,trim=0 0 0 -5]{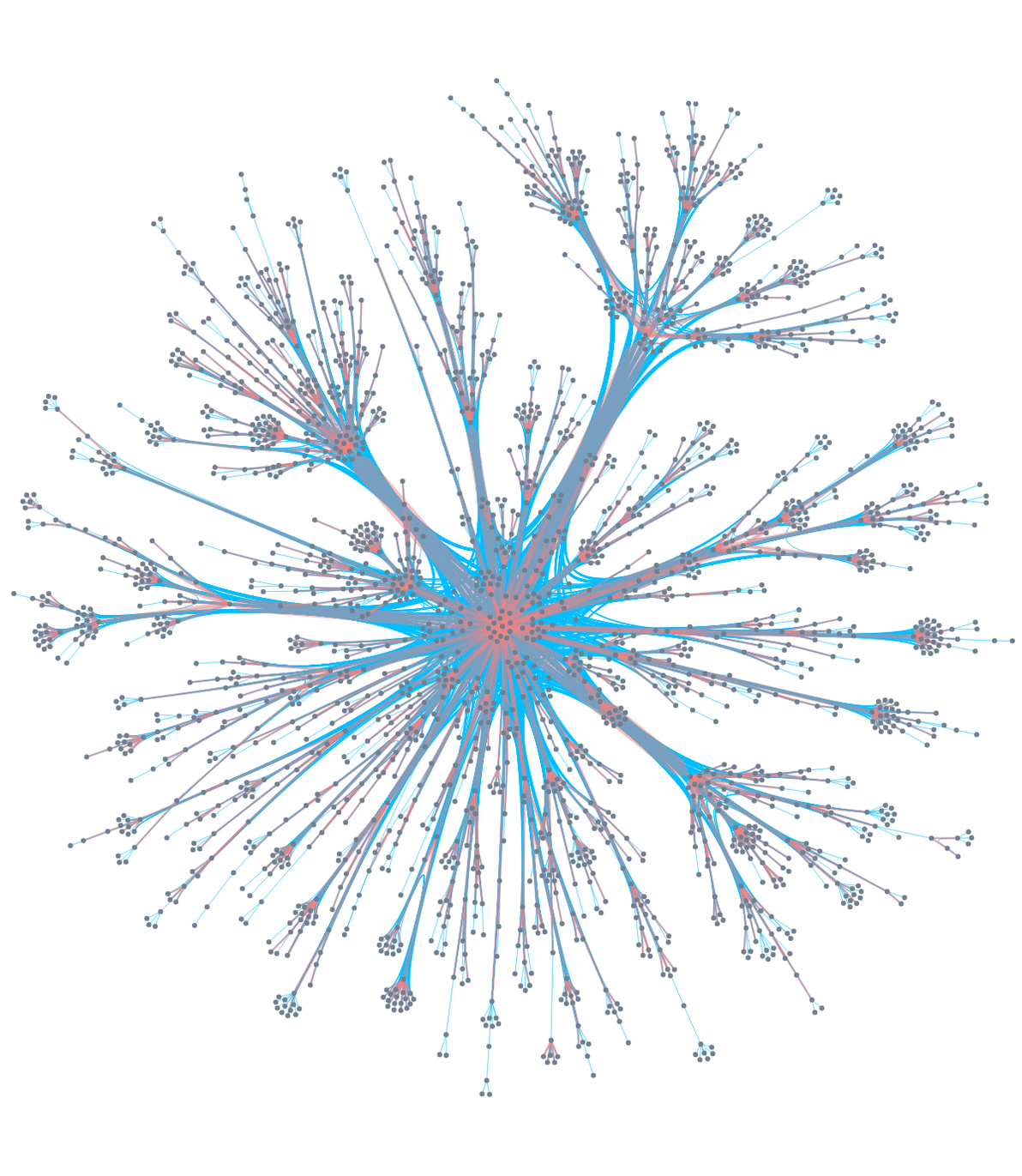}
 \put (-3,95) {\refbigfig{fig:comp:yeast-lstb} \small (\alph{bigfig})}
 \put (0,90) {\small Yeast}
 \put (0,0) {\small LSQT}
\end{overpic} &
\begin{overpic}[width=0.3\textwidth,trim=0 0 0 -5]{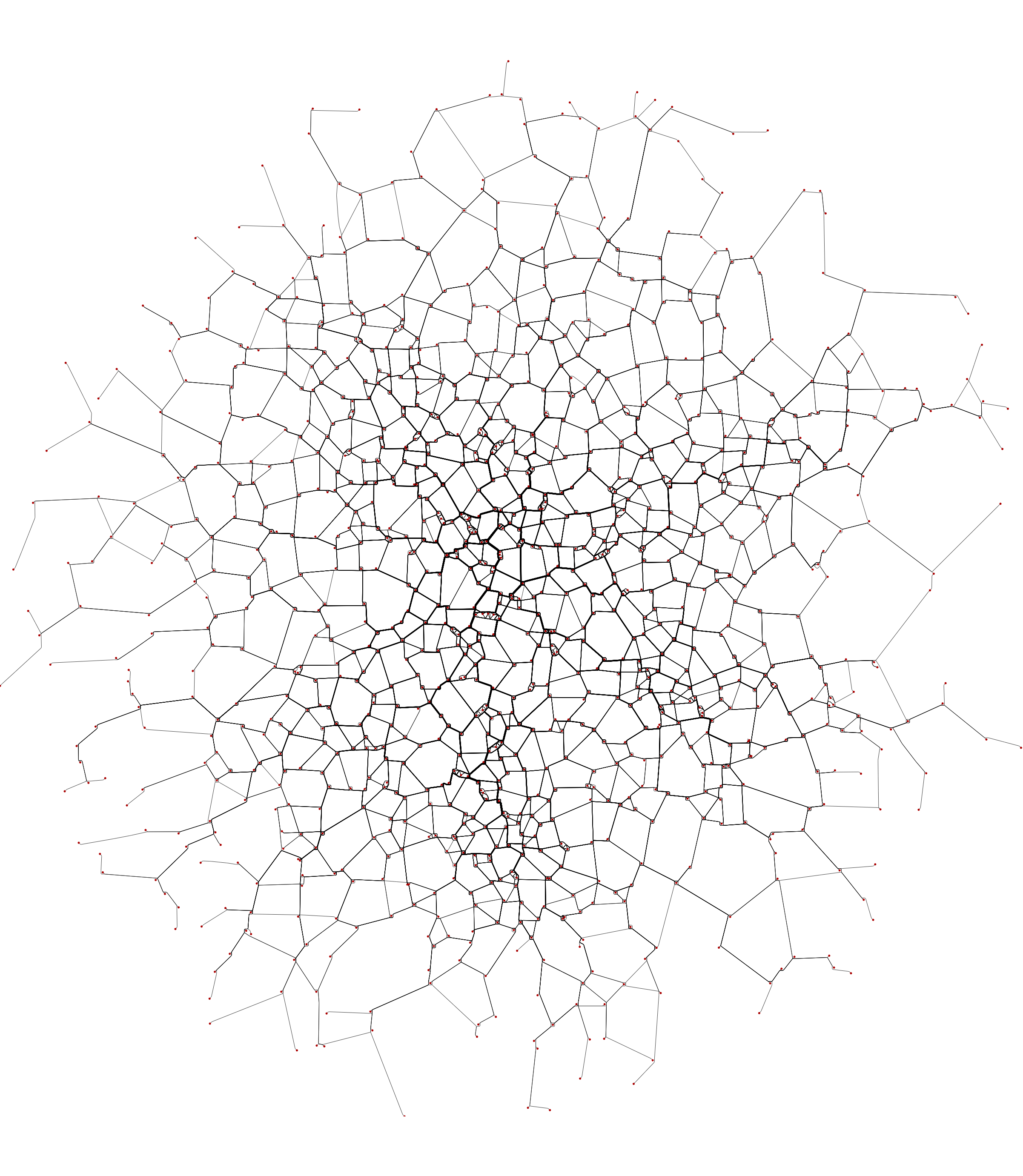}
 \put (-3,95) {\refbigfig{fig:comp:email-bs15} \small (\alph{bigfig})}
 \put (0,90) {\small Email}
 \put (60,95) {\small $|V| = 1133$}
 \put (60,90) {\small $|E|=5451$} 
 \put (0,0) {\small CER~\cite{BS15}}
\end{overpic} &
\begin{overpic}[width=0.3\textwidth,trim=0 0 0 -5]{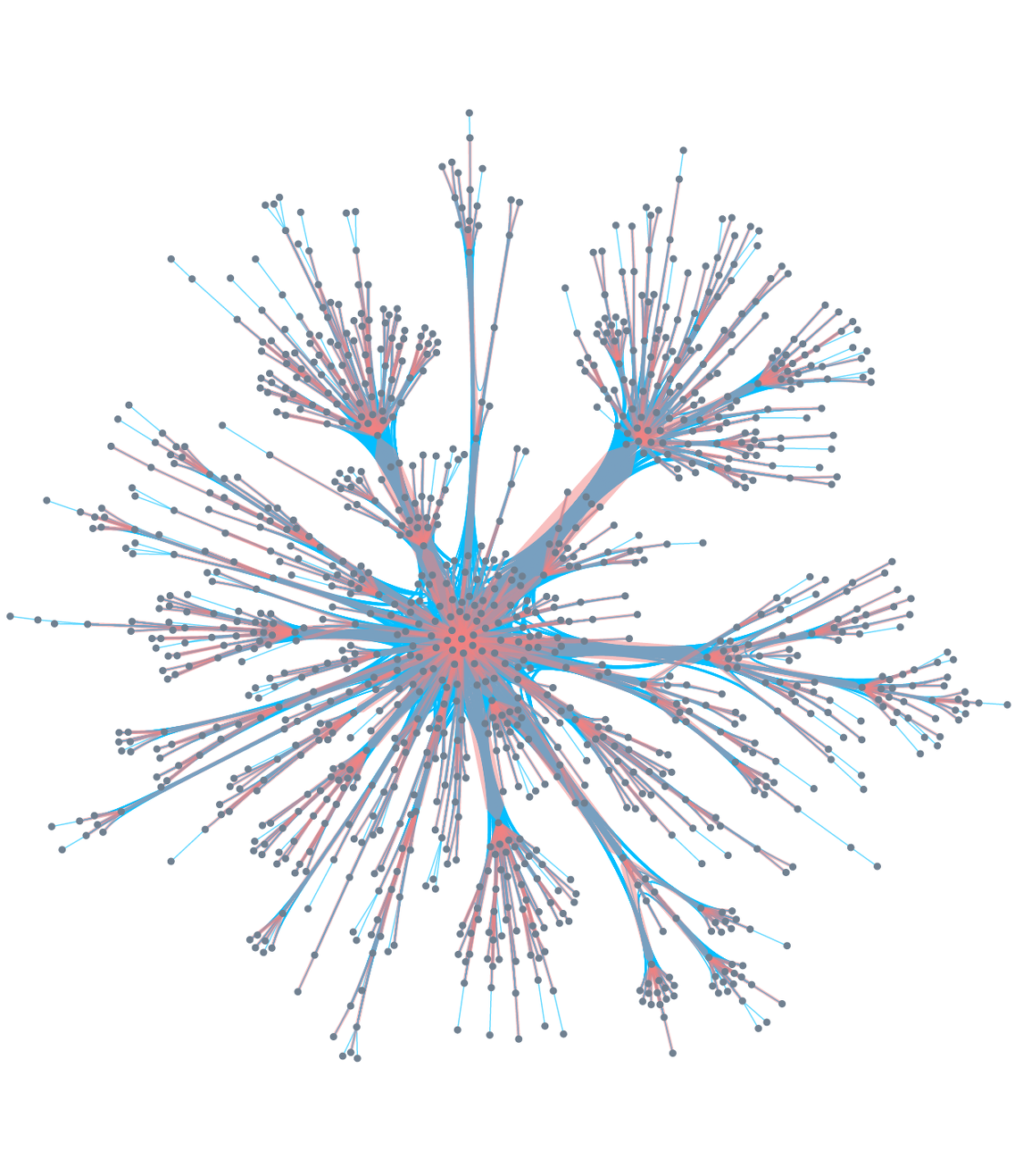}
 \put (-3,95) {\refbigfig{fig:comp:email-lstb} \small (\alph{bigfig})}
 \put (0,90) {\small Email}
 \put (0,0) {\small LSQT}
\end{overpic} \\
\hline
\begin{overpic}[width=0.3\textwidth,trim=0 0 0 0]{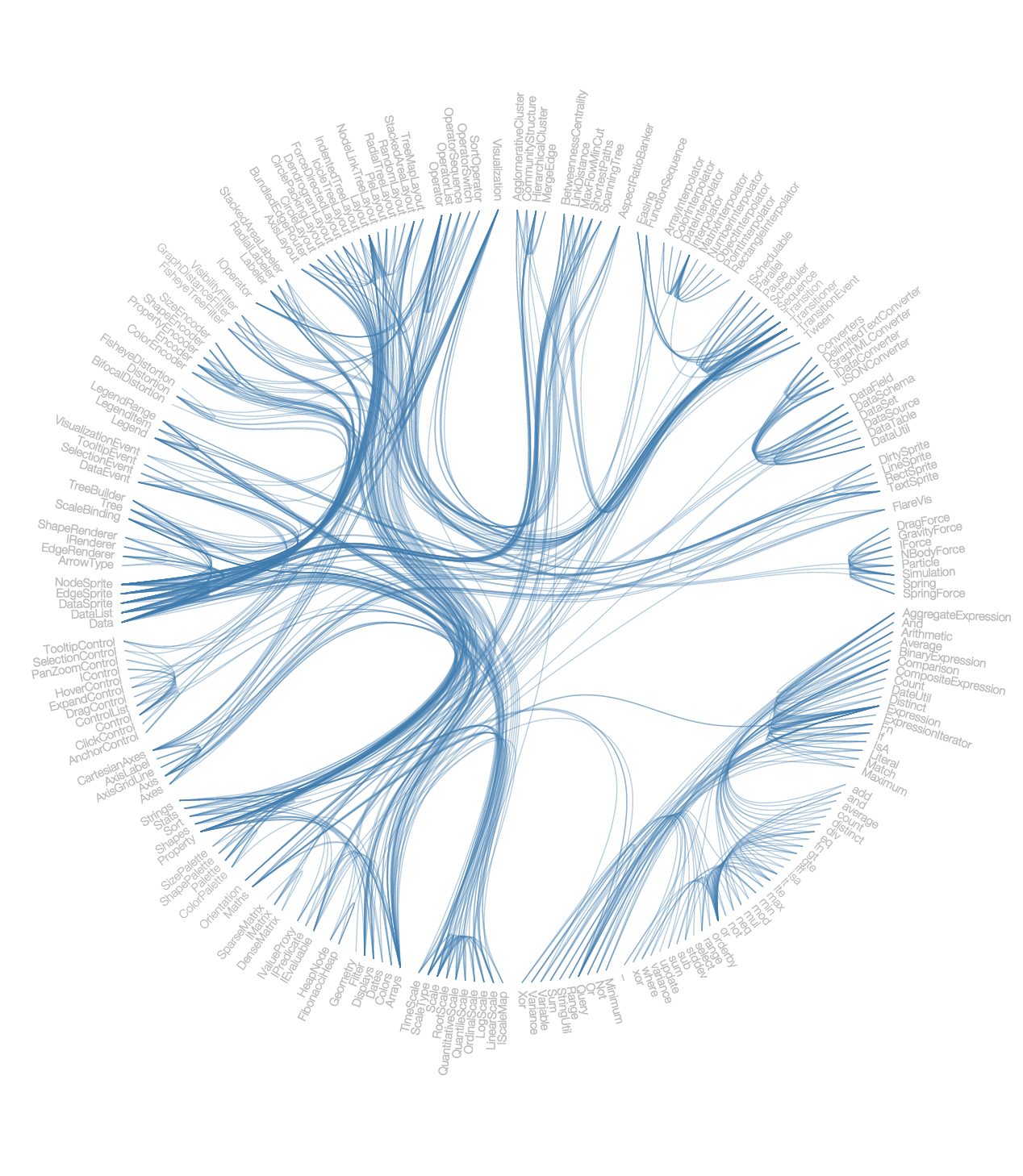}
 \put (-3,95) {\refbigfig{fig:comp:flare-heb} \small (\alph{bigfig})}
 \put (0,90) {\small Flare}
 \put (64,95) {\small $|V| = 220$}
 \put (64,90) {\small $|E|=708$} 
 \put (0,0) {\small HEB~\cite{H06} (D3 implementation \cite{BOH11})}
\end{overpic} &
\begin{overpic}[width=0.3\textwidth,trim=0 0 0 0]{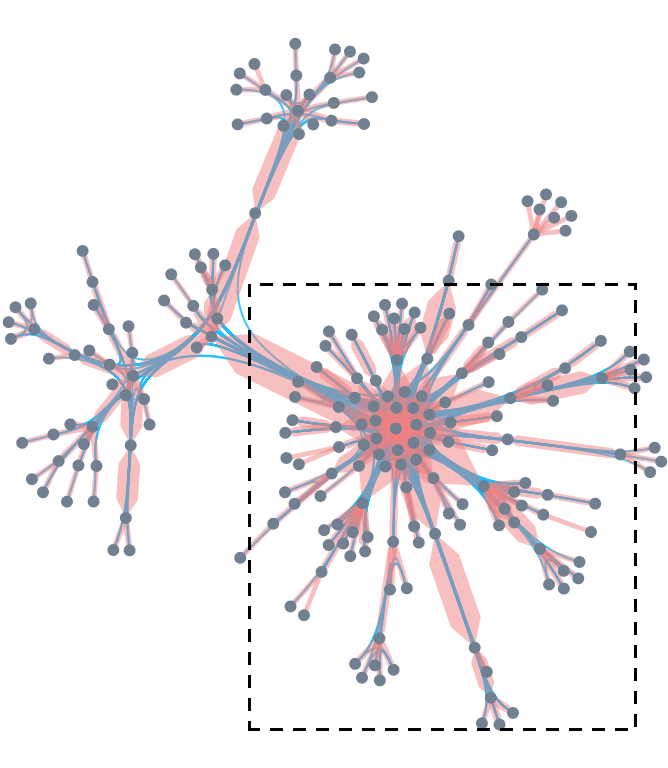}
 \put (-3,95) {\refbigfig{fig:comp:flare-lstb} \small (\alph{bigfig})}
 \put (0,90) {\small Flare}
 \put (0,0) {\small LSQT}
\end{overpic} &
\begin{overpic}[width=0.3\textwidth,trim=0 0 0 -3]{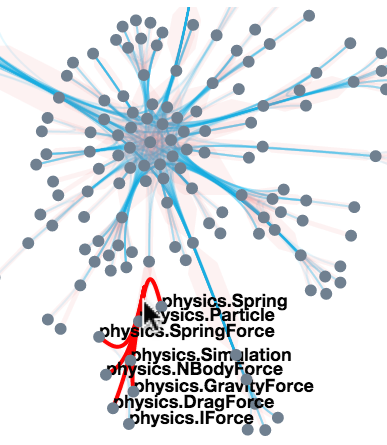}
 \put (-3,95) {\refbigfig{fig:comp:flare-lstb-zoom} \small (\alph{bigfig})}
 \put (0,90) {\small Flare}
 \put (0,0) {\small LSQT}
\end{overpic} \\
\hline
\end{tabular}
  \caption{Comparison between previous methods and LSQT.}
  \label{fig:comp}
\end{figure*}

In this section, we qualitatively compare visual encodings computed with LSQT to encodings of the same data from five previous bundling and layout methods, as shown in  \autoref{fig:comp}.  Four of the methods are layout-first, and we directly show images from those previous papers:  KDEEB~\cite{HET12}, CER~\cite{BS15}, MINGLE~\cite{GHNS11}, SBEB~\cite{EHPC11}. For the fifth method,  the layout-agnostic HEB~\cite{H06}, we created a screenshot using the D3 package that implements this algorithm.

\setlength{\skip\footins}{20pt}
\footnotetext[1]{\scriptsize \url{https://gist.github.com/mbostock/1044242}}
\footnotetext[2]{\scriptsize \url{https://github.com/upphiminn/d3.ForceBundle/tree/master/example/bundling_data}}
\footnotetext[3]{\scriptsize Courtesy of A.\ Telea \cite{HET12}}
\footnotetext[4]{\scriptsize \url{http://yifanhu.net/GALLERY/GRAPHS/}}

The visual results of LSQT are layouts that closely follow the spanning tree backbones. The Wiki and Yeast datasets are examples where previous methods also lead to roughly tree-like structure in the bundled graphs. The Wiki dataset is shown in Figure \hyperref[fig:comp]{\ref*{fig:comp}\ref*{fig:comp:wiki-kdeeb}} with a SBEB layout and in Figure \hyperref[fig:comp]{\ref*{fig:comp}\ref*{fig:comp:wiki-lstb}} with LSQT, and the Yeast dataset is shown in \hyperref[fig:comp]{\ref*{fig:comp}\ref*{fig:comp:yeast-mingle}} with a MINGLE layout and in Figure \hyperref[fig:comp]{\ref*{fig:comp}\ref*{fig:comp:yeast-lstb}} with LSQT. For both datasets, there is a clear core section that is more dense than the branching structure in  periphery in both layouts, so the high-level gist of the structure is qualitatively similar. For the lower-level detail, the LSQT layout provides more visual separation for the leaves of the tree at the periphery, and the SBEB and MINGLE layouts provides more emphasis on the interconnections within the core segment. 

The Email dataset is shown in Figure 
%the LSQT layouts show clear clusters of vertices, as well as disparities in their sizes. 
\hyperref[fig:comp]{\ref*{fig:comp}\ref*{fig:comp:email-bs15}} with a CER layout and Figure \hyperref[fig:comp]{\ref*{fig:comp}\ref*{fig:comp:email-lstb}} with a LSQT layout.  The CER layout emphasizes mesh-like structure, in contrast to the tree-like emphasis of LSQT. The question of which layout is a more appropriate method would depend on the task at hand. Finally, the Flare dataset is shown in Figure \hyperref[fig:comp]{\ref*{fig:comp}\ref*{fig:comp:flare-heb}} with a HEB layout. Figure \hyperref[fig:comp]{\ref*{fig:comp}\ref*{fig:comp:flare-lstb}} shows it with LSTB, and Figure \hyperref[fig:comp]{\ref*{fig:comp}\ref*{fig:comp:flare-lstb-zoom}} is a zoom where the one-hop neighbors of a target node are highlighted with red edges and labels for the directly adjacent nodes. This dataset contains a software package hierarchy where edges represent imports of one package from another. While the low-stretch tree used as a routing graph is not the original package hierarchy, we can see from the package names that similar packages are still clustered together, and hovering over edge bundles makes it easy to see which packages call each other. 

\section{Discussion and Future Work}

LSQT offers promising performance guarantees compared with existing edge-bundling methods. Our simple, unoptimized proof-of-concept implementation of LSQT was able to handle fairly large datasets, up to 7,000 nodes and 100,000 edges, and achieved competitive performance with the fastest previous method. An interesting area for future work would be to adapt our proposed algorithm to exploit GPU parallelism to achieve even better performance.  

LSQT shares some of the strengths and weaknesses of the KDEEB~\cite{HET12} method through the choice to aggressively de-clutter the graph rather than preserve the gist of an exisiting layout; they argue for considering bundle separation as a useful way to assess the visual quality of the resulting layout, and LSQT is also successful by this measure. To go further, our technique enables novel and interesting visual inspection of topological graph structure, rather than geometric structure. While a single LSQT view may be useful for exploratory tasks, we also see potential utility for side-by-side views where a LSQT view is combined with other views showing geometric structure created using other techniques. Visualizations comparing topological and geometric approaches to bundling have not been explicitly investigated in past work, and could inspire new exploratory user tasks. 
%Our method  is similar to KDEEB in that  the same look and feel questions as KDEEB, but similar to their evaluation, we find success in our decluttering and bundling strength. 

% The LSQT segmentation algorithm takes an input graph, and outputs a set of bundles that map between edges and ordered list of segments. Although a few previous techniques do support certain kinds of interaction with bundles, we see the potential of offering many new ways to interact with bundled graphs based on the availability of a data structure that maps these explicitly computed bundles to the original edges.

We call for more layout-agnostic approaches to edge bundling. Surprisingly, the idea of building bundles based on logical topology rather than a previously computed geometric layout has been neglected by all of the subsequent work following the original HEB paper\cite{H06}. While a layout-agnostic approach is not suitable for all tasks, it is well suited for many topology-centered questions.
Previous graph-oriented task taxonomies~\cite{lee2006task} do distinguish between tasks that require understanding topological structure and those that focus on the geometric structure of a particular layout.

%In this vein, future studies might investigate characterizing tasks that require viewing topological versus geometric structure in graphs. Comparing them directly may lead to novel discovery, and we encourage researchers to investigate the value of these kinds of comparative idioms.

 It will be exciting future work to conduct a formal user evaluation of LSQT, possibly in concert with identifying specific user tasks that would benefit the most from this approach. Since LSQT is particularly flexible with regards to layout selection, it could be used for exploratory visual inspection when the choice of layout is not clear in advance, perhaps due to a lack of knowledge about the tasks. 

More broadly, we hope to see more future work that addresses the potential of quasi-trees for both layout and bundling, and that 
%\me{this will be impacted by SBEB paper} Finally, our computation of bundles as a "first class" data abstraction offers new potential for users to think about logical topology in visualization.
%\subsection{Future Work: Visualization}
considers using low-stretch trees in other graph visualization contexts. 
% on low-stretch quasi-trees will serve as a new strategy for hairball graphs, as we demonstrate their strength for aggressive decluttering of graph displays. 

% We also hope that our implementation will inspire other researchers to explore possibilities with bundles as first class data abstractions. For instance, highlighting bundles or remainder edges as a perceptually discriminable foreground layer may increase their utility in visualization tasks. 

%\subsection{Future Work: Algorithms}

At the core of the LSTQ algorithm, alternative approaches to sparsifiers might bear fruit as better routing graphs for edge bundling. New approaches such as Kolla et al.~\cite{KMST10} compute \emph{ultrasparsifiers}, which have $n+o(n)$ edges. Additional work by Lee and Sun~\cite{LS15} improves upon the prior sparsification result by Batson et al.~\cite{BSS12} with a near-linear time algorithm. Both of these approaches could potentially be used to improve the speed and quality of LSQT.

\section{Conclusion}
We have introduced low-stretch trees as a mathematical formalism that is useful in a graph drawing context, and used them as the basis for the new LSQT algorithm that handles both layout and edge bundling with a quasi-tree approach that emphasizes a spanning tree extracted from the graph. We argue that quasi-tree methods are appropriate for a broader class of problems than previously understood, due to the remarkable properties of low-stretch trees that capture with very minimal distortion the underlying structure of graphs that do not appear to be tree-like at first glance. 
%In this paper, we show how quasi-trees have been under-appreciated, and make a case for why they should be studied further for their aggressive de-cluttering ability in hairball graphs. We present LSQT, a novel edge bundling technique that uses low-stretch quasi-trees as a convenient and sophisticated tool for improving quasi-tree implementations.
While previous bundling methods are layout-first, we introduce a way to use the topological features of the graph in order to compute a low-stretch tree, which we then use to route edges without relying on any previously computed geometric layout. 
%This means zero reliance on any input graph layout or hierarchical structure. 
Our bundling method is fast and simple to implement, and provides algorithmic support for sophisticated visual encodings and interactivity.

% or


\ifCLASSOPTIONcompsoc
  \section*{Acknowledgments}
\else
  \section*{Acknowledgment}
\fi

The authors would like to thank Michelle Borkin, Matt Brehmer, Giuseppe Carenini, Anamaria Crisan, Kimberly Dextras-Romagnino, Johanna Fulda, Zipeng Liu, Michael Oppermann, and Emily Hindalong for  discussion and feedback on this work.

\ifCLASSOPTIONcaptionsoff
  \newpage
\fi



%\bibliographystyle{IEEEtran}
\bibliographystyle{abbrv}
\bibliography{refs.bib}

\begin{IEEEbiography}[{\includegraphics[width=1in,height=1.25in,clip,keepaspectratio]{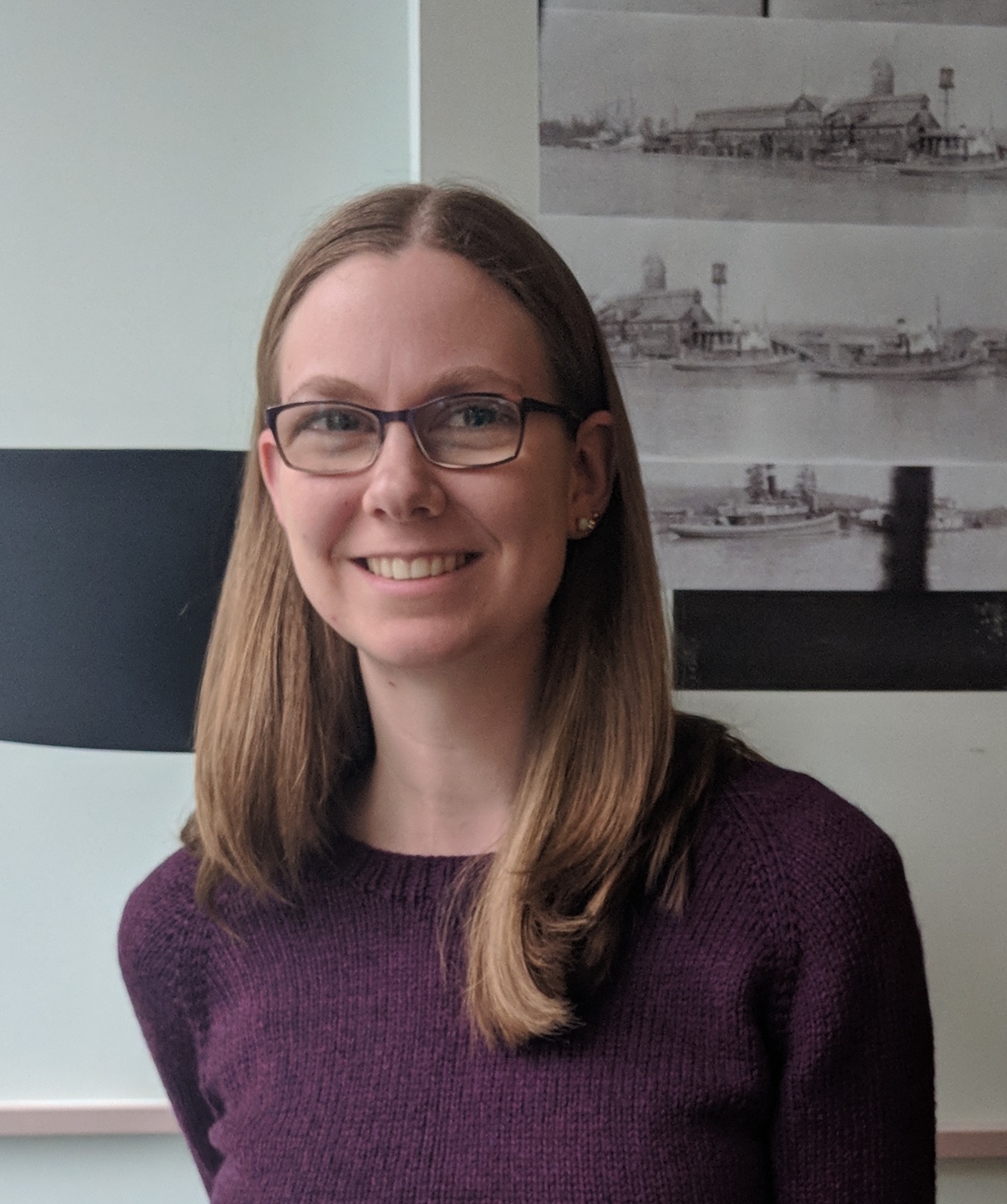}}]{Rebecca Vandenberg}
received a B.Sc.\ (Hons.) in Computer Science from the University of Victoria before commencing studies at the University of British Columbia. She was supervised by Nick Harvey and received her M.Sc.\ in 2015. Since then, she has been working as a Software Development Engineer at Amazon in Vancouver.
\end{IEEEbiography}

\begin{IEEEbiography}[{\includegraphics[width=1in,height=1.25in,clip,keepaspectratio]{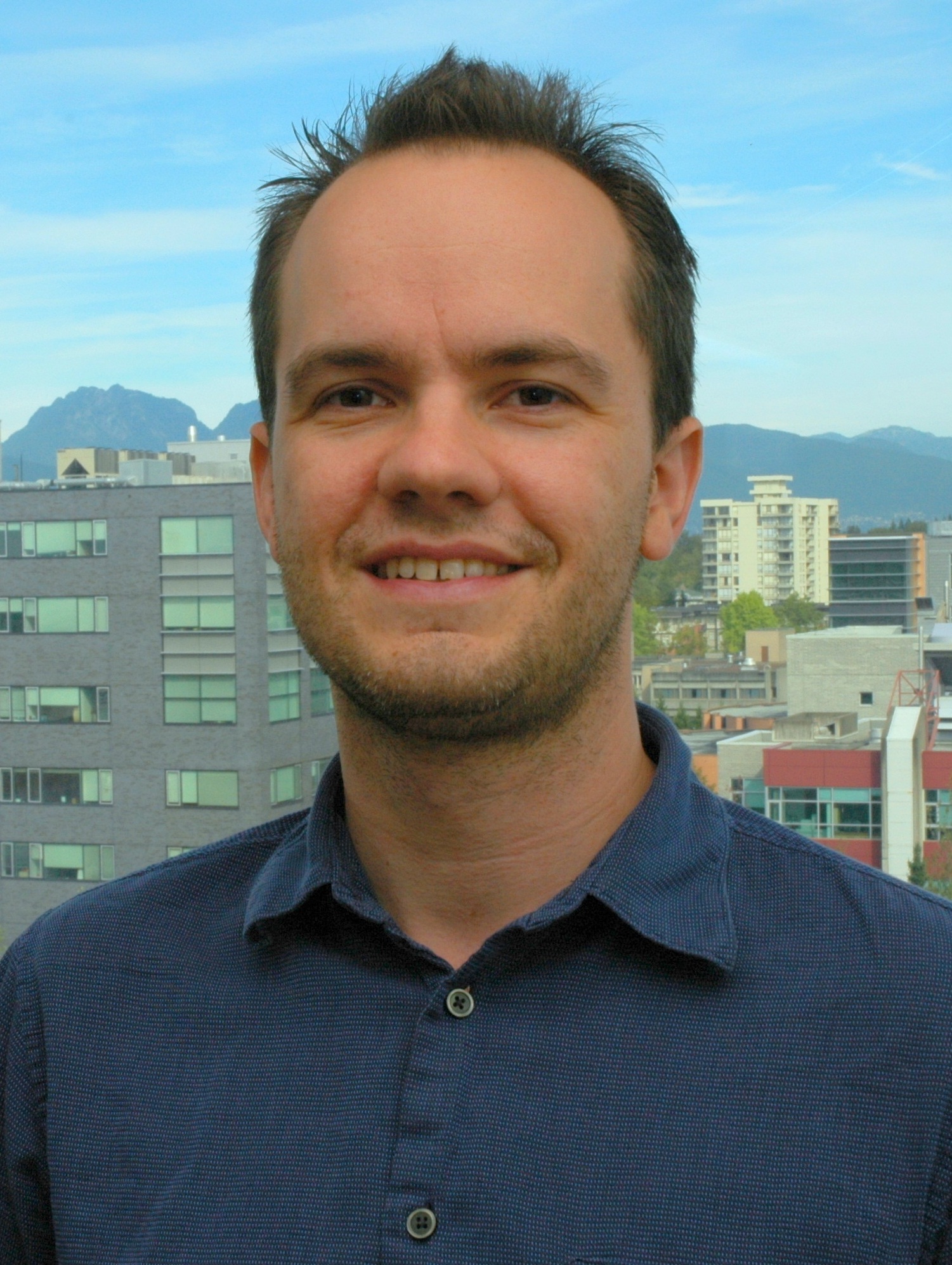}}]{Nicholas Harvey}
is an associate professor at the University of British Columbia
and holds a Canada Research Chair in Algorithm Design.
He received his PhD from MIT in 2008, and a Sloan Research Fellowship in 2013.
His research involves randomized algorithms, optimization and learning theory.
\end{IEEEbiography}

\begin{IEEEbiography}[{\includegraphics[width=1in,height=1.25in,clip,keepaspectratio]{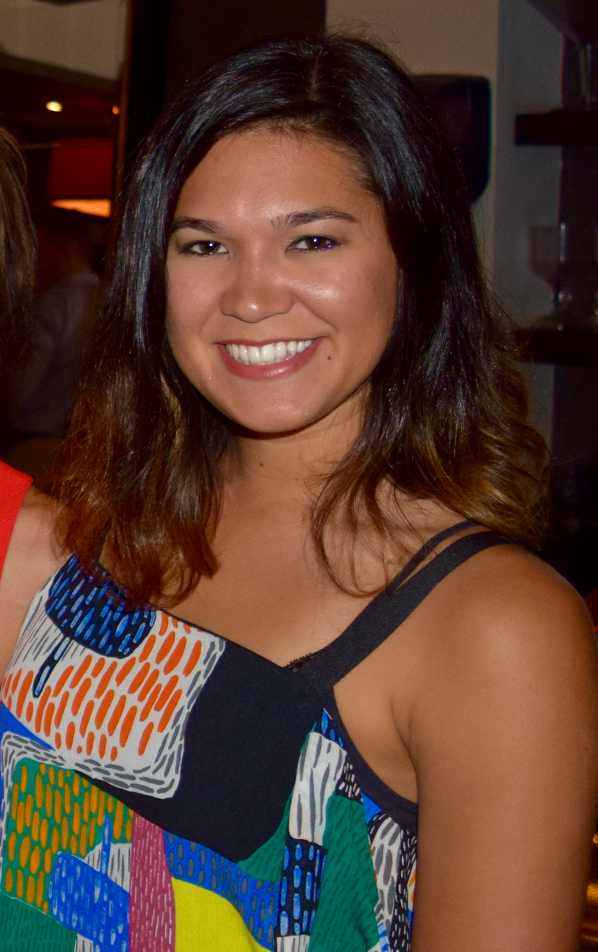}}]{Madison Elliott}
is a PhD student in Cognitive Psychology at the University of British Columbia. She holds an MA in Clinical Psychology from Towson University. Her research investigates the perception of information visualization. Her interests are primarily focused on models of visual attention and feature selection, as well as behavioral research methods.
\end{IEEEbiography}

% \begin{IEEEbiographynophoto}{XXX}

%\newpage

\begin{IEEEbiography}[{\includegraphics[width=1in,height=1.25in,clip,keepaspectratio]{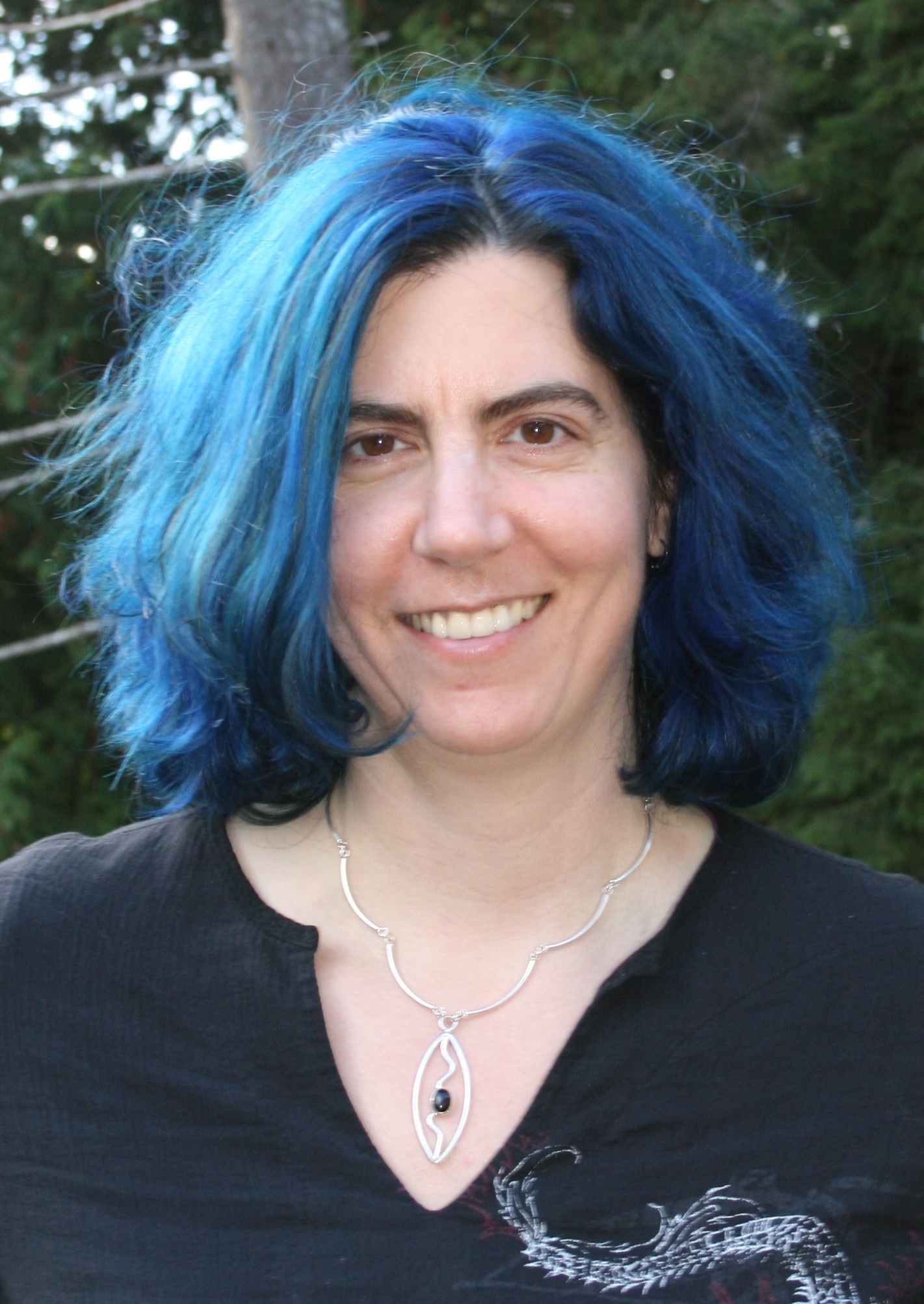}}]{Tamara Munzner}
is a professor at the University of British Columbia, and holds a PhD from Stanford from 2000. She has co-chaired InfoVis and EuroVis, her book \textsl{Visualization Analysis and Design} appeared in 2014, and she received the IEEE VGTC Visualization Technical Achievement Award in 2015. She has worked on visualization projects in a broad range of application domains, including genomics, geometric topology, computational linguistics, system administration, web log analysis, and journalism.
\end{IEEEbiography}


%\vfill


\end{document}